\documentclass[pdflatex,sn-mathphys]{sn-jnl}

\usepackage{stackengine}

\jyear{2022}%

\theoremstyle{thmstyleone}%
\theoremstyle{thmstyletwo}%

\theoremstyle{thmstylethree}%

\begin{document}

\title[Leading effect for wind turbine wake models]{Leading effect for wind turbine wake models}


\author*[1]{\fnm{Ingrid} \sur{Neunaber}}\email{ingrid.neunaber@ntnu.no}

\author[2]{\fnm{Michael} \sur{H\"olling}}\email{michael.hoelling@uni-oldenburg.de}

\author[3]{\fnm{Mart\'in} \sur{Obligado}}\email{Martin.Obligado@univ-grenoble-alpes.fr}

\affil[1]{\orgdiv{Department of Energy and Process Engineering}, \orgname{Norwegian University of Science and Technology}, \orgaddress{\street{Street}, \city{Trondheim}, \postcode{100190}, \country{Norway}}}


\affil[2]{\orgdiv{University of Oldenburg, ForWind}, \orgname{University of Oldenburg}, \orgaddress{\street{Street}, \city{Oldenburg}, \postcode{610101}, \country{Germany}}}

\affil[3]{\orgdiv{Grenoble INP, CNRS, LEGI}, \orgname{Université Grenoble Alpes}, \orgaddress{\street{} \city{Grenoble}, \postcode{38400}, \country{France}}}

\abstract{As wind energy expands worldwide, the demand of reliable, fast, cost-efficient wind turbine wake models is growing. This is a significant challenge as wind turbines face various  inflow conditions, that include turbulence, inhomogeneities/instationarities and upstream wakes. In consequence, an enormous number of engineering models, each one based on different physical concepts, has been proposed. The majority focuses on the far wake where the mean velocity recovers and turbulence decays after it built up. 
We argue that the most important, or the leading, parameter for wake modeling is the length scale of a virtual origin. Testing different models from the literature for data sets from laboratory wind turbines and multi-megawatt turbines obtained by LiDAR, we find that all models perform significantly better when such a virtual origin is added. Our results can therefore be used for a yet missing definition of a near wake zone.
}




\maketitle

\section{Introduction}\label{sec1}

Wind is considered as an important renewable energy source for a future sustainable energy supply of our society \cite{IPCC, Verdolini2021, GIELEN201938}. For economic reasons, wind turbines, more technically speaking, wind energy converters, are mainly build in large farm configurations. For the energy yield, for planning, as well as for operation, the interaction of wind turbines by means of wind-reducing wakes that occur downstream of all operating wind turbines is a major concern. A profound knowledge of the wind speed reduction by wind turbines standing in front of another is essential. 
In consequence, in the wind energy research community, the understanding of the velocity deficit downstream of a wind turbine is one of the most attended topics, as can be seen for example by the number of competing wake models. 

\begin{figure}
\centering
\includegraphics[width=\linewidth]{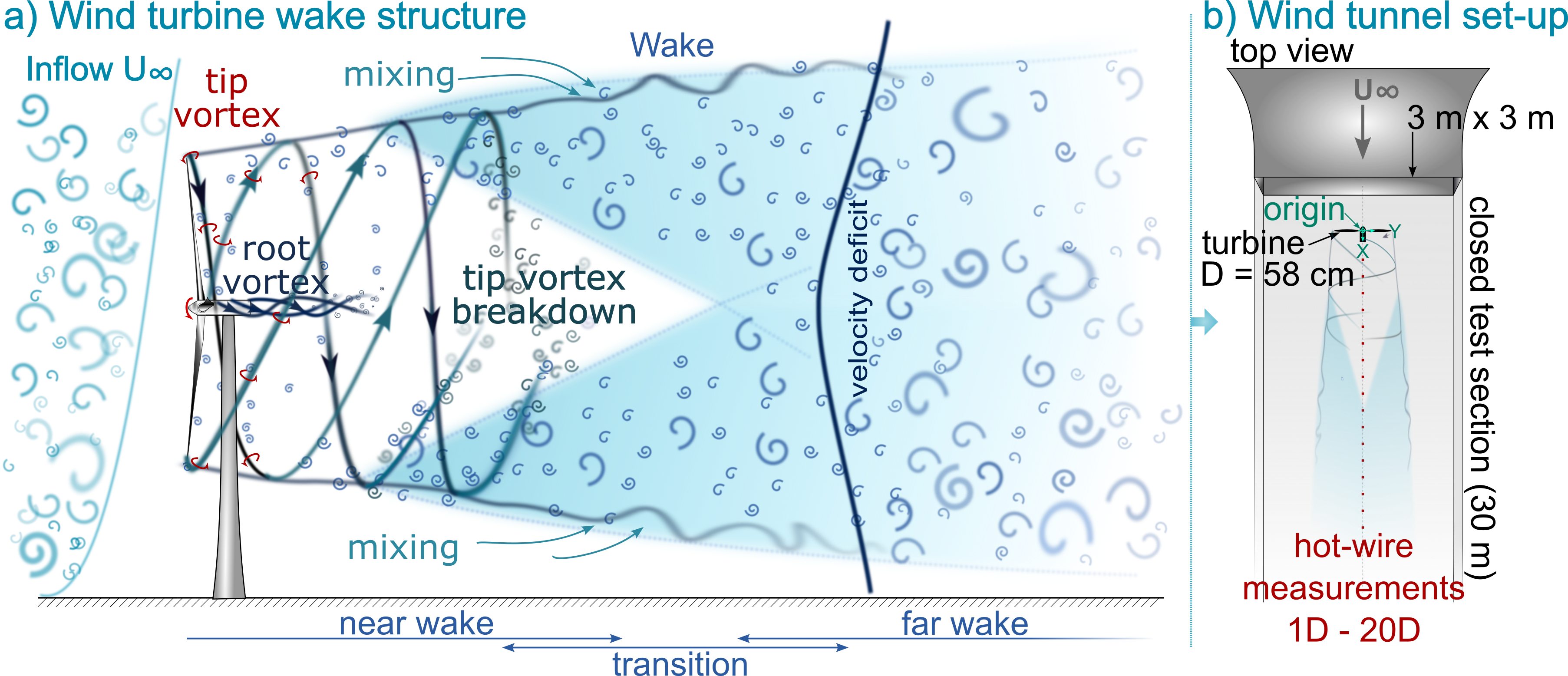}
\caption{a) Schematic of the turbulent wake of a wind turbine, showing the near and far wakes. b) Sketch of the experimental setup used for the wind tunnel tests in this work. \label{fig:Wake}}
\end{figure}

The wake of a wind turbine is classically divided into two major regions, namely the \textit{near wake} and the \textit{far wake} (e.g. \cite{VERMEER2003467,Sanderse09aerodynamicsof}; while other authors mention up to four regions \cite{Neunaber2020,Houtin2021}). In the near wake, the tip and root vortices that are shed from the rotor blades, and the vortices shed from the tower and nacelle dominate the wake (figure \ref{fig:Wake}a). These structures decay and the shear layers surrounding the wake expand while turbulence builds up. Once they meet, around 2-4 rotor diameters downstream of the turbine, the far wake region starts \cite{porteagel2020wind, stevens2017flow, Neunaber2020_Handbook}. It is characterized by a recovery of the velocity in the wake, an approximately Gaussian velocity deficit \cite{porteagel2020wind}, and decaying turbulence that was shown to have strong features of homogeneous, isotropic turbulence \cite{Neunaber2020}. The evolution of the wake depends heavily on the ambient conditions (e.g. \cite{Wu2012, AUBRUN20131, Lundquist2020,Neunaber2020}).

It is the far wake that the vast majority of wind turbine wake models account for \cite{GOCMEN2016752, porteagel2020wind, stevens2017flow}. Indeed, this is a range where simplifying mathematical assumptions can be made, such as self-similarity, the universality of some statistics, etc. However, when applying these wind turbine wake models to wake measurements from field campaigns or wind tunnel measurements, very different results are obtained (e.g. \cite{Campagnolo_2019, LOPES2022104840, JEON20151769}). Despite this discrepancy, in this work we show, by means of data analysis, that these models still capture, in principle, the decay of the velocity deficits in the recovering wakes quite well. 

With the present work we want to draw attention to another important and not well investigated effect, that is, the near wake region directly in the wind shadow of the wind turbine's rotor. While it has been shown that the behavior of the far wake is, in principle, independent both of the inflow turbulence conditions, the turbine operational state and even the wake generating object (i.e., a wind turbine and a static porous disc produce similar far wakes), in the near wake region, turbulence builds up, depending on the inflow conditions etc., until a developed turbulent state is reached from where the decay takes place. We show that it is the not well understood near wake behavior which leads to major differences in common wake models, e.g. \cite{vanderLaan2022_pre}. To overcome this problem, we present a new quantity, a virtual origin, as a length scale for the near wake. This quantity is well known in the field of turbulence, e.g. bluff body wakes \cite{dairay15, obligado2016} and grid-generated turbulent flows \cite{mazellier&vassilicos2010,HearstLavoie,Sinhuber2015}. Based on analyzed data we show evidence that this length parameter is the leading term to improve the wake models and, thus, the description of the wind deficit even in quite far distances behind a wind turbine. 

To justify our claim, we first briefly discuss features of the wake behind a wind turbine as well as experiments performed by us in a wind tunnel with small models of wind turbines (figure \ref{fig:Wake}b). Our results of the measurements with different inflow conditions together with Light Detection And Ranging (LiDAR) measurements from the literature of the wakes behind real multi-megawatt wind turbines are compared with several different standard wake models. Finally we work out that the most important or the leading parameter for wake modeling is the length scale of a virtual origin.

\section{Wind turbine wake modeling}

The evolution of the mean velocity deficit is the most studied quantity in the wake of a wind turbine, followed by the added turbulence intensity and the entrainment of turbulent kinetic energy (e.g. \cite{stevens2017flow, porteagel2020wind, Neunaber2020_Handbook}). Analytical wind turbine wake models are often derived from governing equations and basic assumptions, typically including conservation of mass and momentum. Their aim is a description of the evolution of the mean velocity deficit, and in some cases the turbulence intensity, in the wake based on available quantities such as the inflow averaged streamwise velocity $U_\infty$, the inflow averaged turbulence intensity $TI=\sigma/U_\infty$ (where $\sigma$ denotes the standard deviation of the streamwise velocity) and the thrust coefficient of the wind turbine $c_T$ that indicates the amount of energy converted by the wind turbine. Among these wake models, only few were adopted and are frequently used (for a more complete overview about different wind turbine wake models, the reader is referred to e.g. \cite{porteagel2020wind,GOCMEN2016752, HEGAZY2022457, Schmidt2020}). A brief description of the models that will be checked in this work, their main hypotheses and governing equations can be found in the supplementary material. We remark that, while they all provide a functional form for the centreline velocity deficit, some allow to obtain also the wake width via different relations (momentum conservation, a hypothesis on the wake profile shape, etc.). Such models will therefore provide the main relevant information for the wake's averaged quantities. 

Finally, we remark that almost no model considers the existence of a virtual origin. Besides a ``streamwise offset for the reference frame'' that is added by the models proposed (or, rather, updated) in \cite{Schreiber2020} and \cite{larsen2009simple}, the only exception is the Townsend-George model~(see \cite{Townsend,george1989} for the original description and~\cite{dairay2015} for its extension to a non-Kolmogorov energy cascade). In this framework, the virtual origin arises naturally from the governing equations. While this model has originally been developed for bluff bodies, it has recently been shown that the wake of a wind turbine fulfils the requirements necessary to apply it \cite{Neunaber2022}, and several works that apply this phenomenology to wind turbine wakes found a better fit compared to engineering wind turbine wake models \cite{okulov2015wake,stein2019non, Neunaber2021, Neunaber2022}. In the folloing, the virtual origin will therefore be added to all other models as an \textit{ad hoc} assumption to quantify, at first order, the influence of the near wake and the production of turbulence (as it is already the case in the Townsend-George phenomenology and also when studying the evolution of turbulence downstream of different grids, e.g. \cite{mazellier&vassilicos2010, Mora2019,Sinhuber2015}). 

\section{Results}\label{sec2}

The results of this paper are based on data collected in controlled conditions in a wind tunnel and field measurements behind multi-megawatt wind turbines obtained with LiDAR technique (e.g. \cite{LiDAR, Gottschall2020}). Indeed, to justify our claims, field data are the most important data sets, but have the drawback that the precise working mode of the turbine is not always known, nor the exact inflow condition, which is always unsteady so that mean wind velocities and the turbulence intensity may change during measuring period (cf. \cite{Morales2012, Waechter2012, Milan2013}). Such disadvantages can be overcome by controlled experiments with wind turbine models in a wind tunnel. The biggest disadvantage of wind tunnel experiments is the low Reynolds number but it is commonly assumed that the wake behaviour resembles well that of the free field \cite{WHALE1996621}. Here, we use both kinds of data: the field data from the literature mentioned above and wind tunnel data form the set-up described below in section \ref{sec3}. We obtain consistent results for all different data sets. The collection methods used for all sets of data are detailed in section~\ref{sec3}.

\begin{figure}   
    \begin{minipage}[t]{.49\textwidth}
        \centering
        \stackinset{l}{0.1cm}{t}{0.2cm}{\parbox{0.5cm}{(a)}}{
        \includegraphics[width=\textwidth]{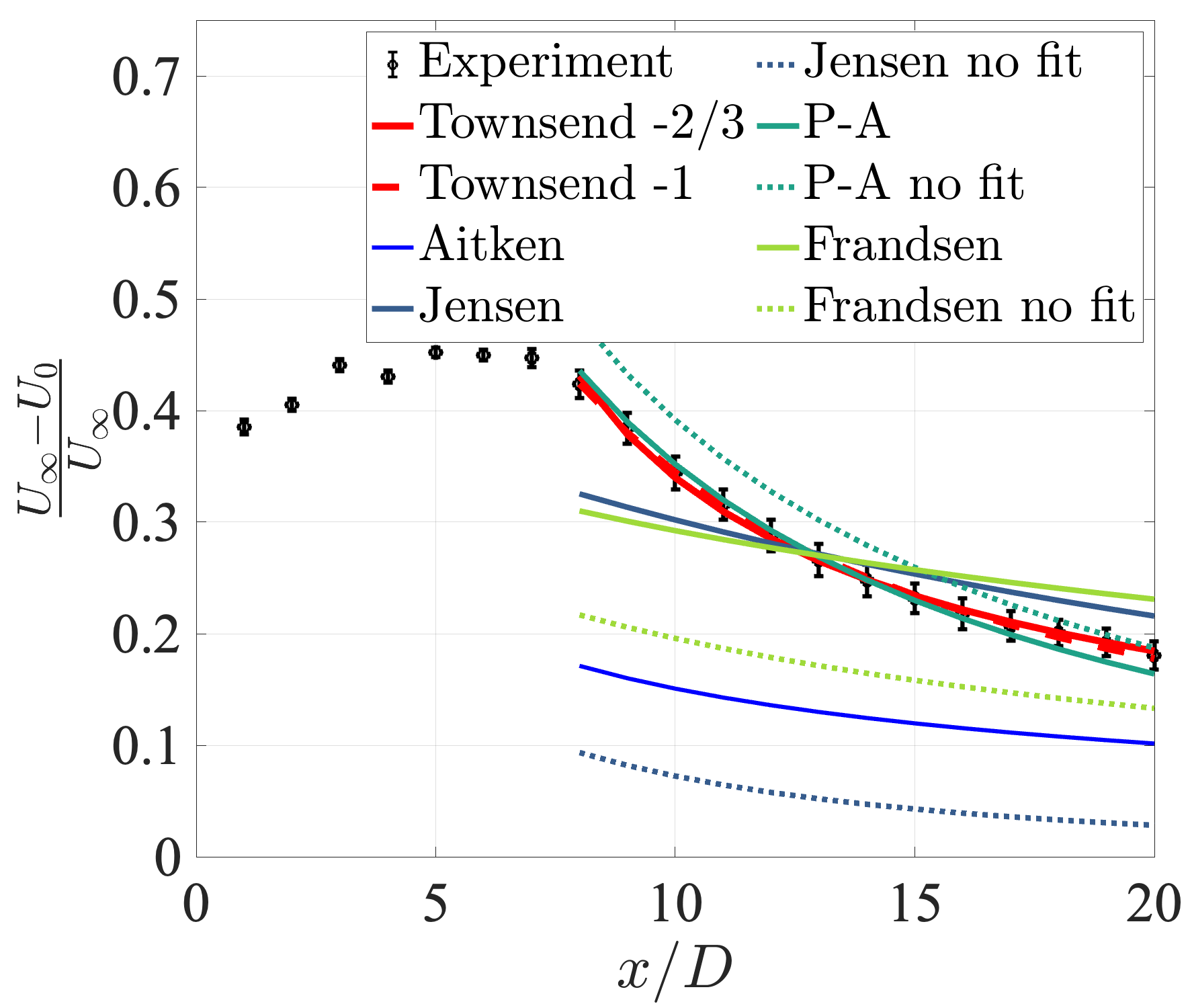}
        }
    \end{minipage}
    \hfill
    \begin{minipage}[t]{.49\textwidth}
        \centering
        \stackinset{l}{0.1cm}{t}{0.2cm}{\parbox{0.5cm}{(b)}}{
        \includegraphics[width=\textwidth]{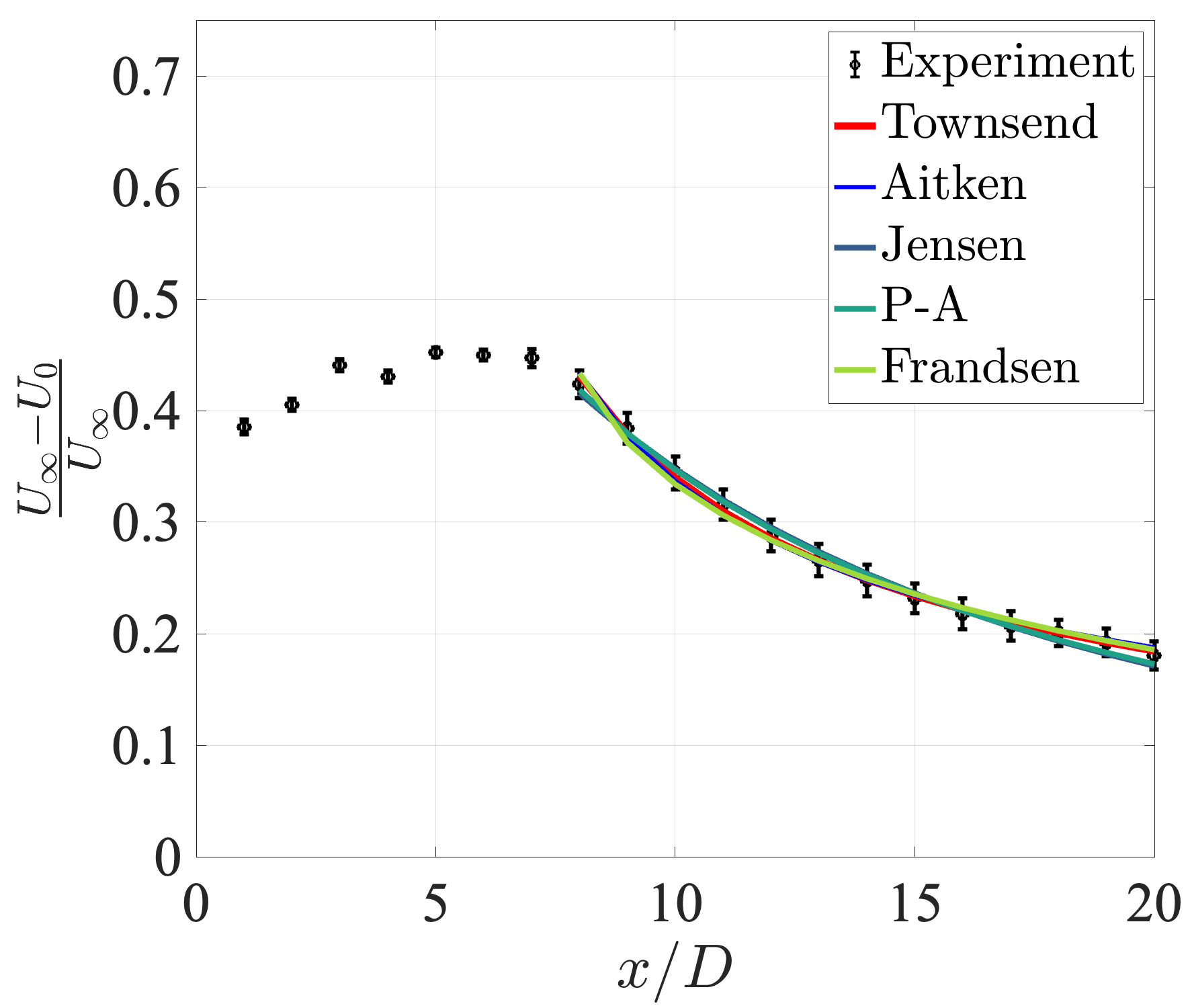}
        }
    \end{minipage}  
    \caption{Downstream evolution of the normalized mean velocity deficit for wind tunnel experiments. Laminar inflow with no virtual origin (a), laminar inflow with virtual origin (b). The legend 'no fit' corresponds to the model used with all parameters extracted from the literature and/or turbine's properties. \label{fig:stream}}
\end{figure}

We start by analysing the data obtained in lab conditions. This data set was measured in the wake of a scaled rotor for both laminar and turbulent inflows.
Figure~\ref{fig:stream} shows the results for a laminar inflow, where the experimental data and the predictions of six different models can be compared. While all models describe the wake with different grades of accuracy (figure \ref{fig:stream}(a)), by only adding a virtual origin they all significantly improve their performance (figure \ref{fig:stream}(b)). Note that the Townsend-George models are used in their native form and thus always include the virtual origin. By just adding a virtual origin they all seem to fit the profile equally well, almost collapsing onto a single curve among them and with the experimental data. The fitting parameters for all figures presented in this work can be found in the supplementary material. Moreover, when changing in this experiment the inflow conditions
to turbulent ones exactly the same trend is obtained
(see supplementary material).

\unskip

\begin{figure}
    \begin{minipage}[t]{.49\textwidth}
        \centering
        \stackinset{l}{0.1cm}{t}{0.2cm}{\parbox{0.5cm}{(a)}}{
        \includegraphics[width=\textwidth]{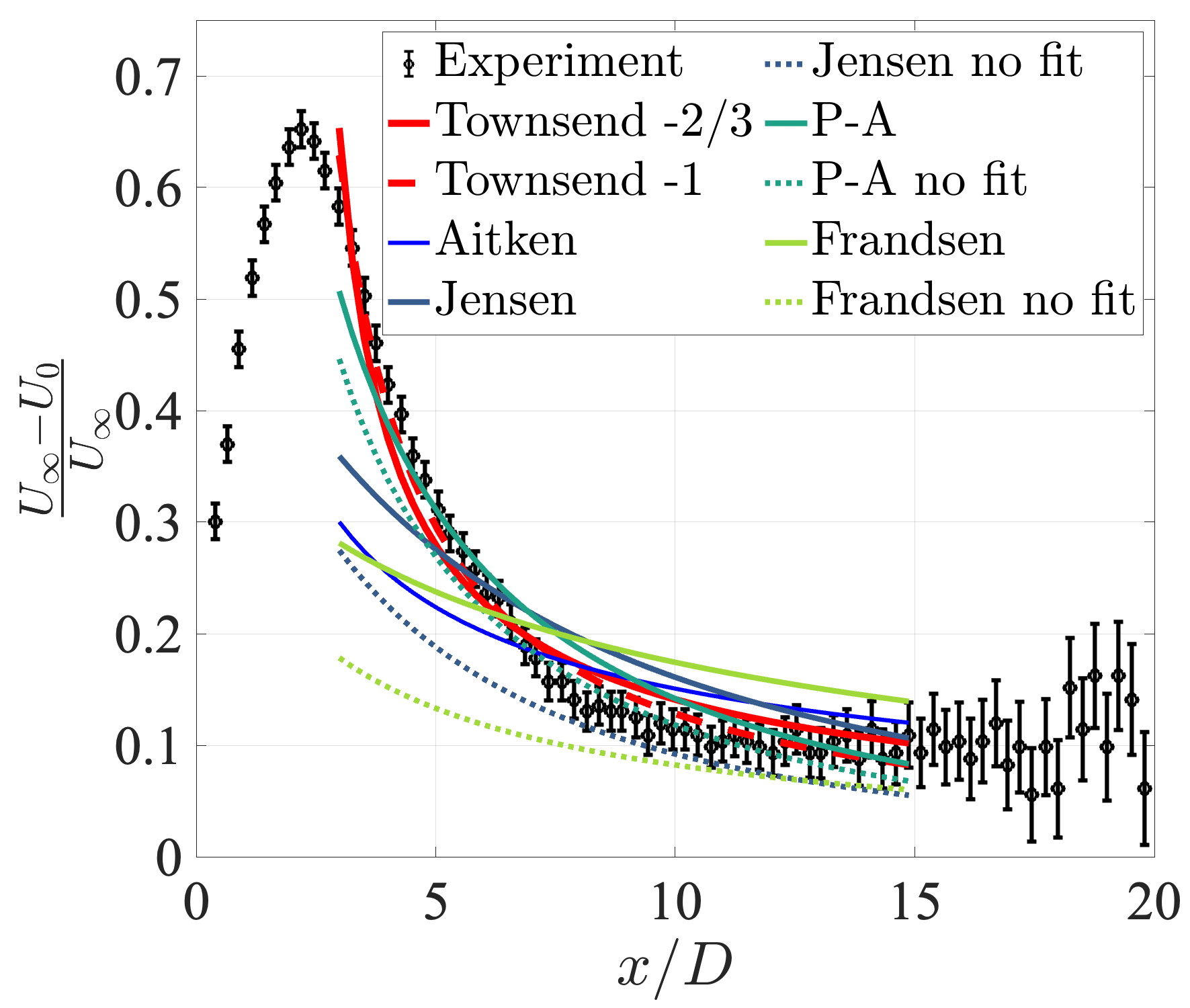}
        }
    \end{minipage}
    \hfill
    \begin{minipage}[t]{.49\textwidth}
        \centering
        \stackinset{l}{0.1cm}{t}{0.2cm}{\parbox{0.5cm}{(b)}}{
        \includegraphics[width=\textwidth]{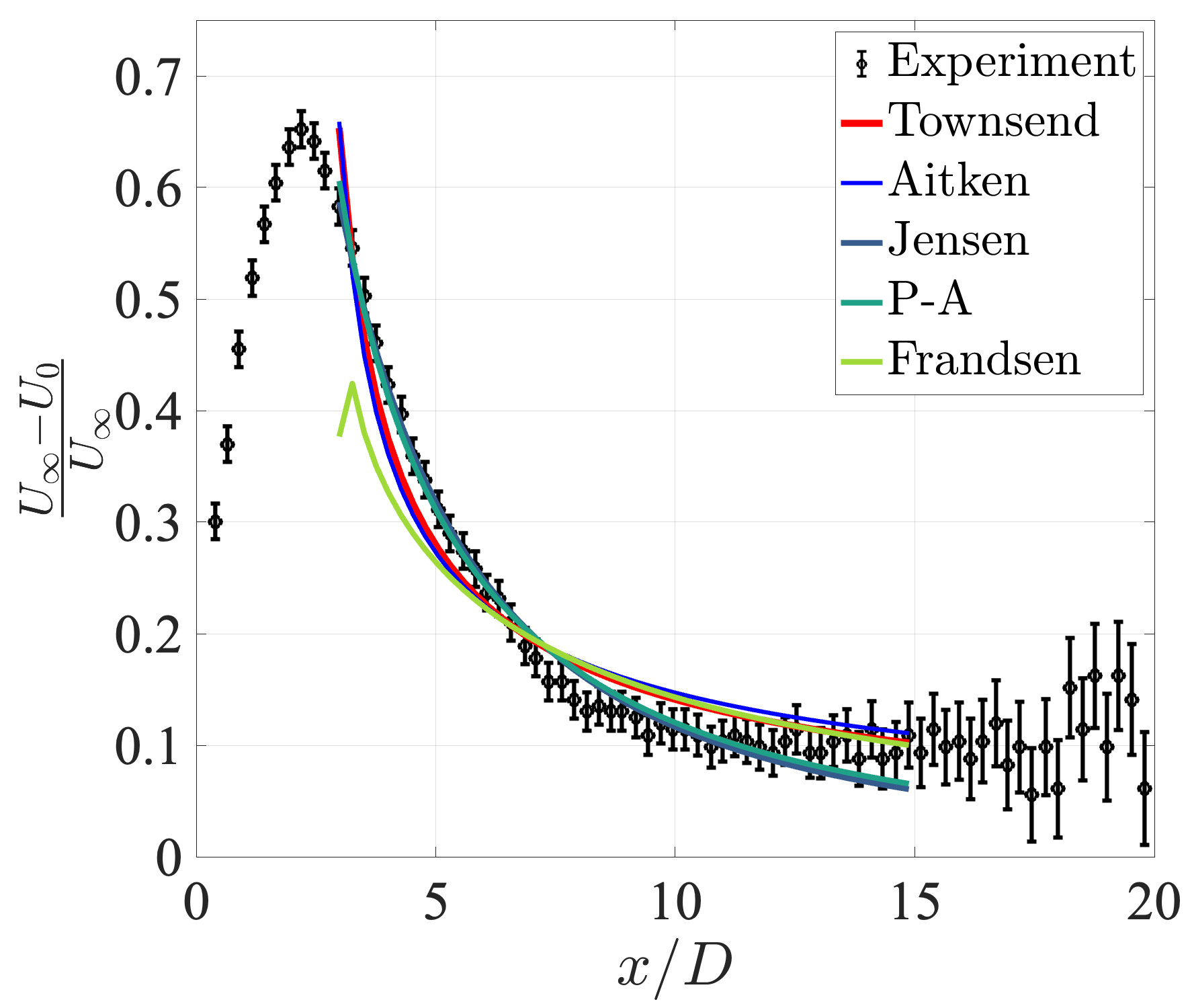}
        }
    \end{minipage}  
    \caption{Downstream evolution of the normalized mean velocity deficit: Data measured by LiDAR with a velocity range of $6-8\:\text{ms}^{-1}$, extracted from \cite{Gallacher2014} with $c_T$ values from \cite{Soesanto2022}. No virtual origin (a), Virtual origin (b). The legend 'no fit' corresponds to the model used with all parameters extracted from the literature and/or turbine's properties. \label{fig:Gallacher}}     
\end{figure}

\begin{figure}
    \begin{minipage}[t]{.49\textwidth}
        \centering
        \stackinset{l}{0.1cm}{t}{0.2cm}{\parbox{0.5cm}{(a)}}{
        \includegraphics[width=\textwidth]{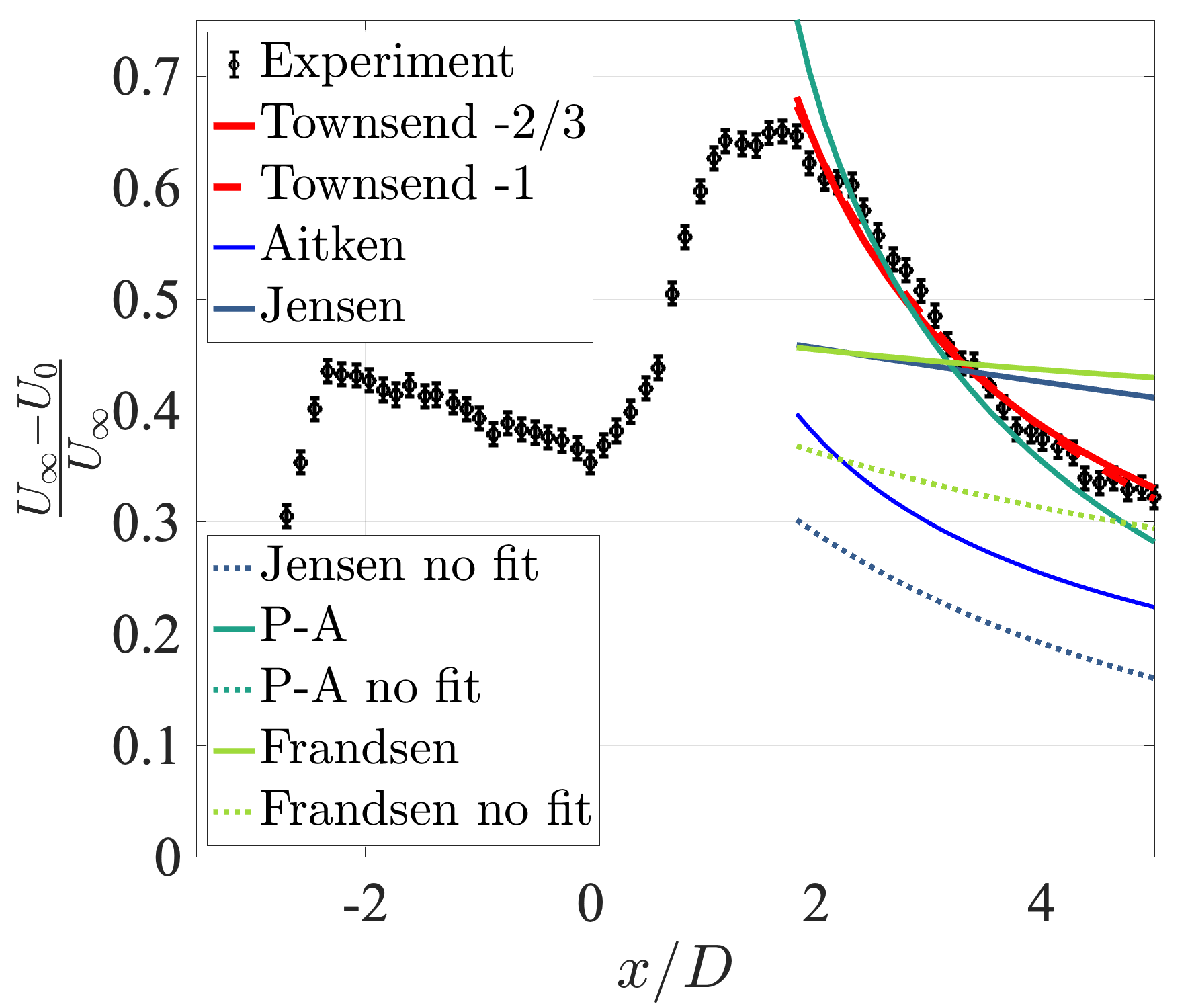}
        }
    \end{minipage}
    \hfill
    \begin{minipage}[t]{.49\textwidth}
        \centering
        \stackinset{l}{0.1cm}{t}{0.2cm}{\parbox{0.5cm}{(b)}}{
        \includegraphics[width=\textwidth]{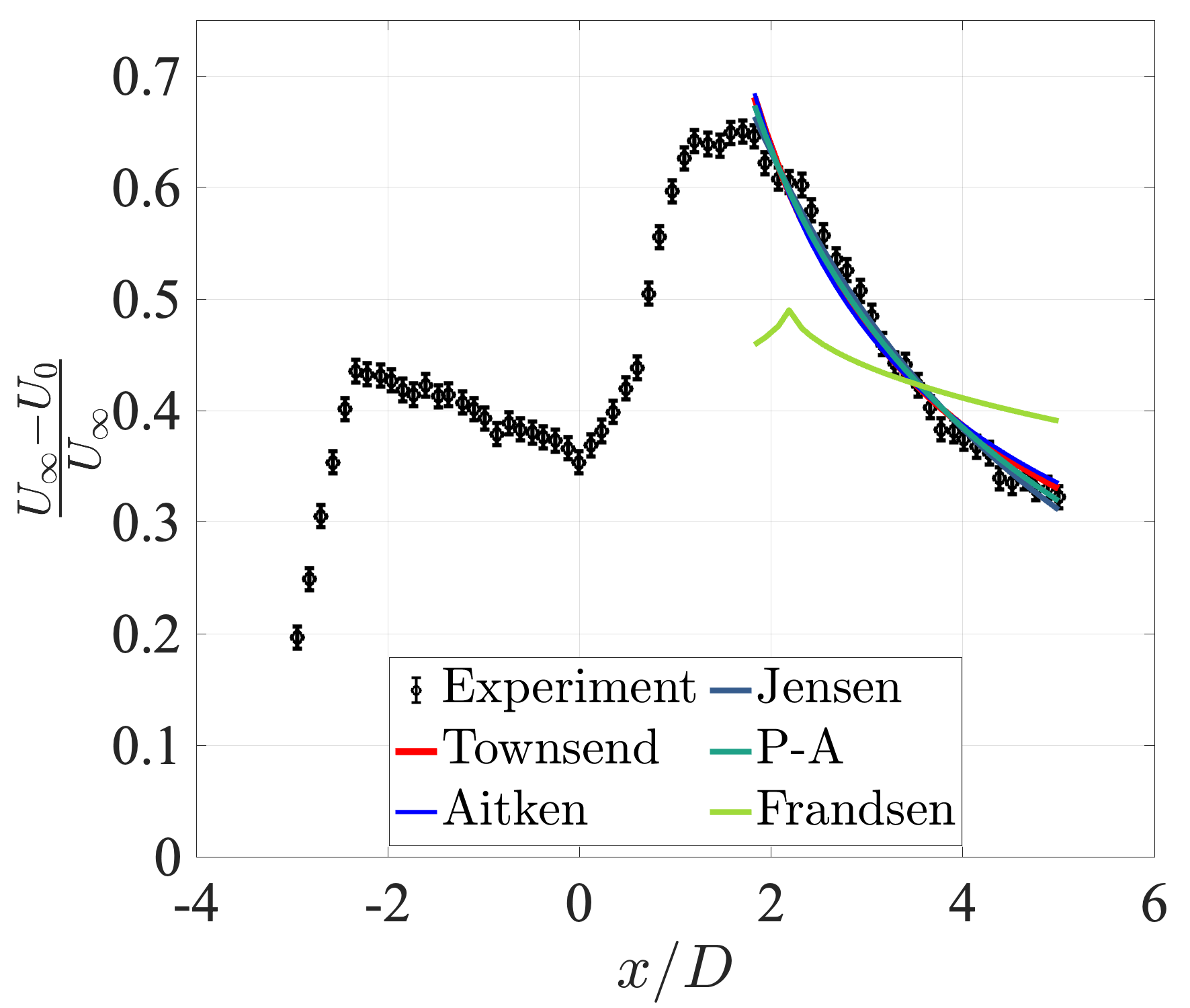}
        }
    \end{minipage}  
    \caption{Downstream evolution of the normalized mean velocity deficit: Data measured by LiDAR, extracted from \cite{HEGAZY2022457}; the wind turbine is in the wake of an upstream turbine. No virtual origin (a), Virtual origin (b).     \label{fig:Smarteole}
}
\end{figure}

We now proceed to verify if a similar improvement is observed when dealing with LiDAR field data extracted from the literature. We check several data sets that include isolated wakes (figure \ref{fig:Gallacher}), and double-wakes (\textit{i.e.}, wakes with an inflow that is the wake of an upstream turbine, figure \ref{fig:Smarteole}). More data sets with different inflow velocities and different degrees of wind turbine interaction can be found in the supplementary material. When a virtual origin is added, they all follow the same trends shown in figures \ref{fig:Gallacher} \& \ref{fig:Smarteole}.
 
\unskip

Field data has larger error bars and shows behaviours that do not fully follow a decaying axisymmetric turbulent wake. For instance, the velocity deficit in figure \ref{fig:Gallacher} tends to a constant value instead of showing the monotonous decrease expected for an isolated wake in its far field region. This shows the difficulties in modelling realistic wind turbine-generated flows in atmospheric conditions, as the resulting flow is actually a combination of different canonical flows (atmospheric boundary layers, wakes, etc.) plus the wake properties may change during data acquisition. In consequence, models tend to perform worse than in wind tunnel conditions. Nevertheless, our main finding is that all models produce good fits by only adding a virtual origin. In consequence, we conclude that by modelling the turbulence production range with the length scale of a virtual origin, all wake models tend to give acceptable results, no matter the inflow or conditions for which data was collected.

\unskip

\section{Methods}\label{sec3}
\subsection{Experimental Setup}
The measurements presented here were carried out in the wind tunnel of the University of Oldenburg. The closed-loop wind tunnel has an inlet of $3\times 3 \:\text{m}^2$ and a test section with a length of $30\:\text{m}$ \cite{kroger2018generation}. The setup is presented in figure \ref{fig:Wake}(b), and the origin of the coordinate system is at the center of the turbine rotor. Experiments were carried out using a three-bladed horizontal axis model wind turbine of type MoWiTO 0.6 (cf. \cite{Schottler2016}) with a control system described in \cite{petrovic2018wind}. It has a rotor diameter of $D=58\:\text{cm}$ and a thrust coefficient of $c_T = 0.70$ for optimal operation conditions at tip speed ratio $TSR \approx6.5$.  The blockage of the turbine with additional support is below 5\%. When operating in the so-called laminar conditions (i.e., with the rotor placed in an empty test section), the turbine is exposed to a low turbulent inflow conditions with a velocity of $U_\infty = 8.3\:\text{ms}^{-1}$ and turbulence intensity of $TI = 0.3\%$ (and therefore a Reynolds number $Re_D=U_\infty D /\nu=3.2\times 10^5$, with $\nu$ the kinematic viscosity of air). The turbulent inflow is generated by means of a passive grid with a velocity of $U_\infty = 6.9\:\text{ms}^{-1}$ and a turbulence intensity of $TI = 2.0\%$ ($Re_D=2.7\times10^5$). To characterize the mean velocity at the centerline of the rotor $U_0$ (i.e., $Y = Z = 0$), a 1d hot-wire was traversed between $x/D = 1$ and $x/D = 20$ in steps of $1D$. It was sampled at $f_s = 20\:\text{kHz}$; a hardware low-pass filter with a cut-off frequency of $f = 10\:\text{kHz}$ was set. The sampling time was of 240s.
\subsection{Field measurements}
Data from two field campaigns carried out using LiDAR is utilized here. This technology is capable of measuring the line of sight velocity at multiple points nearly simultaneously, and the wind speed can be estimated from this by sweeping the azimuth angle at a fixed elevation angle. This type of scan is called plan-position-indicator (PPI) scan. The method is thus suitable to investigate the centerline velocity downstream of a wind turbine. The data sets were extracted using a web plot digitizer \cite{WebPlotDigitizer}.\\
The first data set (cf. figure \ref{fig:Smarteole}) stems from a measurement campaign carried out in a wind farm in the north of France (cf. \cite{TORRESGARCIA20191} and \cite{HEGAZY2022457}). A LiDAR was positioned approximately $1300\:\text{m}$ from the investigated wind turbines at the ground, and the elevation angle of the PPI scans used here is $3.8^\circ$, leading to a variation of the measurement height below $\pm 0.1 D$ in the fit region. The data was filtered according to atmospheric stability (neutral), environmental wind speed ($13\:\text{ms}^{-1}$ at $50\:\text{m}$) and wind direction ($207^\circ$; results for $233^\circ$ and $246^\circ$ are in the supplementary material). The data set presented here shows the wake of a turbine exposed to the wake of an upstream turbine. The turbulence intensity of the inflow and the thrust coefficient are taken from \cite{HEGAZY2022457}. Based on the variations of the data points, error bars were added that indicate fluctuations of $0.01\cdot \frac{U_\infty-U_0}{U_\infty}$.\\
The second data set (cf. \ref{fig:Gallacher}) stems from a campaign carried out at the wind farm Alpha Ventus in the German North Sea using a LiDAR mounted on the nacelle of a wind turbine. The PPI scans have therefore zero elevation angle and were obtained slightly above the centerline. The data was filtered according to wind speed ($(6-8)\:\text{ms}^{-1}$; more data sets are in the supplementary material) and wind direction ($210^\circ - 330^\circ$ for an undisturbed inflow), and the wake centerline velocity was extracted. Error bars represent the standard deviation over all samples normalized by the number of samples. The thrust coefficient was estimated by \cite{Soesanto2022} who also use the data sets.
\section{Discussion}\label{sec12}
By means of a few comprehensive examples, we show that the addition of a virtual origin to widely used engineering models for wind turbine wakes has the capability of improving all these models significantly. This simple but effective addition to wake models is motivated by fundamental turbulence research, and its significance extends to serving as a novel measure of the near wake region that is, contrary to the far wake, not universal. In particular, we show how the decay law of the mean centerline velocity deficit is secondary to the implementation of a virtual origin. In addition, the virtual origin has to be seen as a dynamic correction. The objective being that, once the underlying physics are understood, this length scale can be easily tuned to unsteady operation conditions, i.e., the inflow velocity, inflow turbulence degree and thrust coefficient of the turbine.

With the importance of precise modeling of wind turbine wakes, our findings shift the discussion of wind turbine wakes in a new direction. Therefore, the next step is to understand the physics of the virtual origin that are determined by the flow effects occurring in the near wake.
    
    
    
    
    

\backmatter


\bmhead{Acknowledgments}
Part of this work has been performed during the stay associated to fellowship from the Hanse-Wissenschaftskolleg (HWK Institute for Advanced Study, Delmenhorst, Germany) assigned to MO and IN. The authors also thank the Lille Turbulence Program (LTP) and Joachim Peinke for fruitful discussions.

\end{document}


\title[Supplementary Material: Leading effect for wind turbine wake models]{Supplementary Material: Leading effect for wind turbine wake models}

\section{Fit Parameters}\label{secA1}

\begin{sidewaystable}
\sidewaystablefn%
\begin{center}
\begin{minipage}{\textwidth}
\caption{Fit parameters for all data sets for the Townsend bluff body wake models, and the Jensen, Porté-Agel, Aitken and Frandsen wake models. For the Jensen model, the numbers marked as \textbf{bold} are values from literature (according to \cite{porteagel2020wind}, 0.075 for onshore and 0.04-0.05 for offshore wakes), or close to literature. For the Frandsen model, it is stated that $k_F =\mathcal{O}(10\cdot k_j)$ \cite{porteagel2020wind} for small $c_T$.}\label{tab1a}
\tiny
\begin{tabular*}{\textwidth}{@{\extracolsep{\fill}}lllllrlrlrlrrrrr@{\extracolsep{\fill}}}
\toprule%
                &               &       &                            & \multicolumn{2}{c}{Townsend -2/3} & \multicolumn{2}{c}{Townsend -1} & \multicolumn{2}{c}{Jensen}          & \multicolumn{2}{c}{Porté-Agel} &          \multicolumn{2}{c}{Aitken} &         \multicolumn{2}{c}{Frandsen}          \\
                \cmidrule{5-6}\cmidrule{7-8}\cmidrule{9-10}\cmidrule{11-12}\cmidrule{13-14}\cmidrule{15-16}
 Data set               &  fit             & $c_T$ & $TI$                       & $T_{-2/3}$            & $x_0/D$          & $T_{-1}$           & $x_0/D$         & $k_j$  & $x_0/D$ & $k_{BP}$        & $x_0/D$ & $A$    & $x_0/D$ & $k_f$      & $x_0/D$ \\
                \midrule
                & $k$           &       &                            &                &                  &               &                 & 0.011  & 0       & 0.015      & 0       & 3.756  & 0       & 0.028    & 0       \\
laminar         & $k$, $x_0$ & 0.70   & 0.0012\footnotemark[1] & 1.203          & 3.337            & 3.671         & -0.655          & 0.024  & 7.109   & 0.014      & -1.271  & 2.292  & 4.415   & 0.074    & 7.830    \\
                & $x_0$         &       &                            &                &                  &               &                 & \textbf{0.027}  & 7.500     & $k_j$      & 4.108    & 1      & 6.738   & $3\cdot k_j = 0.081$    & 8.250    \\
                & $k$           &       &                            &                &                  &               &                 & 0.009  & 0       & 0.014      & 0       & 4.423  & 0       & 0.020     & 0       \\
turbulent       & $k$, $x_0$ & 0.70   & 0.02                       & 1.426          & 2.377            & 4.552         & -2.239          & 0.021  & 7.574   & 0.011      & -2.578  & 3.017  & 3.639   & 0.057    & 8.176   \\
                & $x_0$         &       &                            &                &                  &               &                 & \textbf{0.024}  & 8.109   & 0.012\footnotemark[2]      & -2.195  & 1      & 6.894   &$2.5\cdot k_j =  0.060$     & 8.256   \\
                & $k$           &       &                            &                &                  &               &                 & 0.051  & 0       & 0.045      & 0       & 1.213  & 0       & 0.128    & 0       \\
Krishnamurthy   & $k$, $x_0$ & 0.80   & 0.10                        & 0.479          & 1.910             & 0.897         & 1.093           & 0.110   & 2.639   & 0.059      & 0.984   & 0.585  & 2.105   & 0.314    & 2.876   \\
                & $x_0$         &       &                            &                &                  &               &                 & \textbf{0.075}  & 2.012   & 0.042\footnotemark[2]      & -0.070   & 1      & 1.435   & $5\cdot k_j = 0.375$    & 3.876   \\
                & $k$           &       &                            &                &                  &               &                 & 0.009  & 0       & 0.039      & 0       & 2.590   & 0       & 0.006    & 0       \\
Smarteole a T2  & $k$, $x_0$ & 0.74  & 0.18\footnotemark[3]                  & 0.939          & 0.208            & 1.946         & -1.064          & 0.062  & 2.955   & 0.030       & -0.835  & 1.803  & 0.560    & 0.026    & 2.187   \\
                & $x_0$         &       &                            &                &                  &               &                 & \textbf{0.075}  & 2.916   & 0.042\footnotemark[2]      & 0.136   & 1      & 1.272   & $k_j = 0.075$    & 2.322   \\
                & $k$           &       &                            &                &                  &               &                 & 0.022  & 0       & 0.046      & 0       & 0.871  & 0       & 0.070     & 0       \\
Smarteole c T1  & $k$, $x_0$ & 0.53  & 0.11                       & 0.488          & 0.777            & 1.002         & -0.429          & 0.070   & 2.792   & 0.041      & -0.403  & 0.575  & 1.097   & 0.182    & 2.328   \\
                & $x_0$         &       &                            &                &                  &               &                 & \textbf{0.075}  & 2.817   & 0.046\footnotemark[2]      & -0.069  & 1      & -0.270   & $3\cdot k_j = 0.225$    & 2.446   \\
                & $k$           &       &                            &                &                  &               &                 & 0.005  & 0       & 0.047      & 0       & 1.209  & 0       & 0.012    & 0       \\
Smarteole c T2  & $k$, $x_0$ & 0.58  & 0.11                       & 0.891          & -1.262           & 2.138         & -3.280           & 0.041  & 2.822   & 0.023      & -2.826  & 1.524  & -0.688  & 0.048    & 1.972   \\
                & $x_0$         &       &                            &                &                  &               &                 & \textbf{0.075}  & 2.723   & 0.046\footnotemark[2]      & -0.142  & 1      & 0.362   & $ k_j = 0.075$    & 2.184   \\
                & $k$           &       &                            &                &                  &               &                 & 0.040   & 0       & 0.073      & 0       & 0.660   & 0       & 0.126    & 0       \\
Smarteole d T1  & $k$, $x_0$ & 0.58  & 0.10                        & 0.493          & -0.094           & 0.952         & -1.038          & 0.083  & 1.769   & 0.048      & -0.969  & 0.603  & 0.149   & 0.223    & 1.312   \\
                & $x_0$         &       &                            &                &                  &               &                 & \textbf{0.075}  & 1.751   & 0.042\footnotemark[2]      & -1.331  & 1      & -0.712  & $3\cdot k_j = 0.225$    & 1.312   \\
                & $k$           &       &                            &                &                  &               &                 & 0.024  & 0       & 0.058      & 0       & 0.924  & 0       & 0.061    & 0       \\
Smarteole d T2  & $k$, $x_0$ & 0.61  & 0.10                        & 0.568          & 0.148            & 1.161         & -1.099          & 0.073  & 1.973   & 0.041      & -1.067  & 0.749  & 0.501   & 0.209    & 2.018   \\
                & $x_0$         &       &                            &                &                  &               &                 & \textbf{0.075}  & 1.996   & 0.042\footnotemark[2]      & -0.994  & 1      & -0.184  &$3\cdot k_j =  0.225$    & 2.080    \\
                & $k$           &       &                            &                &                  &               &                 & 0.045  & 0       & 0.037      & 0       & 1.423  & 0       & 0.117    & 0       \\
Gallacher 6-8   & $k$, $x_0$ & 0.82  & 0.10                        & 0.555          & 2.200              & 1.130          & 1.190            & 0.088  & 3.040    & 0.046      & 1.059   & 0.725  & 2.441   & 0.247    & 3.259   \\
                & $x_0$         &       &                            &                &                  &               &                 & \textbf{0.075}  & 2.803   & 0.042\footnotemark[2]      & 0.753   & 1      & 2.106   & $5\cdot k_j = 0.375$    & 3.508   \\
                & $k$           &       &                            &                &                  &               &                 & 0.040   & 0       & 0.034      & 0       & 1.589  & 0       & 0.104    & 0       \\
Gallacher 8-10  & $k$, $x_0$ & 0.81  & 0.10                        & 0.609          & 2.108            & 1.265         & 1.026           & 0.081  & 3.231   & 0.042      & 0.941   & 0.842  & 2.367   & 0.229    & 3.477   \\
                & $x_0$         &       &                            &                &                  &               &                 & \textbf{0.075}  & 3.145   & 0.042\footnotemark[2]      & 0.932   & 1      & 2.198   & $5\cdot k_j = 0.375$    & 3.749   \\
                & $k$           &       &                            &                &                  &               &                 & 0.039  & 0       & 0.035      & 0       & 1.542  & 0       & 0.103    & 0       \\
Gallacher 10-12 & $k$, $x_0$ & 0.79  & 0.10                        & 0.625          & 1.910             & 1.347         & 0.683           & 0.073  & 3.088   & 0.039      & 0.508   & 0.866  & 2.220    & 0.221    & 3.476   \\
                & $x_0$         &       &                            &                &                  &               &                 & \textbf{0.075}  & 3.126   & 0.042\footnotemark[2]      & 0.779   & 1      & 2.049   & $5\cdot k_j = 0.375$    & 3.745  
\\
\botrule
\end{tabular*}
\footnotetext[1]{$k_j$ used for the Porté-Agel model since the equation for estimating $k_{BP}$ is not valid for very low TIs.}
\footnotetext[2]{Calculated with the equation from \cite{porteagel2020wind}: $0.38\cdot TI+0.004$.}
\footnotetext[3]{For the calculation of $k$ according to the Porté-Agel model, $TI=0.1$ has been used here to reach convergence of the fit.}
\end{minipage}
\end{center}
\end{sidewaystable}

\section{Wind turbine wake models}\label{secA2}
\subsection{Townsend-George model for free shear flows}
From the assumptions of self-similarity of the one-point turbulence quantities and high-Reynolds number turbulence decaying in equilibrium as postulated in \cite{kolmogorov1941}, \cite{townsend1949} and \cite{george1989} derive the recovery of the normalized mean velocity deficit of the far wake of a bluff body
\begin{equation} \label{eq:EQ_U}
\frac{\Delta U}{U_\infty} = T_{-2/3} \cdot \left((x-x_0)\right)^{-2/3}.
\end{equation}
Here, $\Delta U=U_\infty-U_0$ denotes the centerline velocity deficit at downstream position $x$, $U_\infty$ is the inflow velocity, $T_{-2/3}$ is a scaling parameter that we will tune here and $x_0$ is a virtual origin. In addition, this model allows for the  calculation of the wake width $\delta$,
\begin{equation} \label{eq:EQ_d}
\delta (x) = B_{-2/3} \cdot  \left((x-x_0)\right)^{1/3},
\end{equation}
where $B_{-2/3}$ gives the wake growth.\\
However, it has also been shown by \cite{george1989} that a second scaling exists,

\begin{equation} \label{eq:NEQ_U}
\frac{\Delta U}{U_\infty} = T_{-1} \cdot \left((x-x_0)\right)^{-1},
\end{equation}
\begin{equation} \label{eq:NEQ_d}
\delta (x) = B_{-1}\cdot \left((x-x_0)\right)^{1/2}.
\end{equation}


\subsection{Jensen wake model}
The Jensen wake model is a pioneering wake model proposed by N.O. Jensen in 1983 \cite{jensen1983note}. It assumes a top-hat shaped velocity deficit and is derived by means of conservation of mass. The normalized velocity deficit is given by
\begin{equation}
    \frac{\Delta U}{U_\infty}=\frac{1-\sqrt{1-c_T}}{(1+2k_j x/D)^2}.
\end{equation}
$c_T$ is the turbine's thrust coefficient. A linear expansion of the wake is assumed, and it is described by the wake growth rate $k_j$. Onshore, $k_j=0.075$ is suggested whereas offshore, where the wake recovery is slower due to lower turbulence levels, $k_j=0.04$ or $k_j=0.05$ is assumed \cite{Barthelmie2009}; \cite{GOCMEN2016752}.\\
Here, we tune the model by a) treating $k_j$ as a fitting parameter, and b) treating $k_j$ as a fitting parameter while also replacing $x$ with $x-x_0$ (and therefore $x_0$ is also a fitting parameter), thus adding a virtual origin. Also, we show in section \ref{secA4} how adding a virtual origin while using literature values for $k_j$ improves the modeling.\\
While there are several improvements to the original model, see e.g. \cite{Campagnolo_2019} for an overview, in this study, we use the original version.

\subsection{Porté-Agel wake model}
From conservation of mass and momentum, \cite{bastankhah2014new} derived a model for the description of the three-dimensional far wake of a wind turbine assuming a Gaussian shape of the normalized velocity deficit,
\begin{align}\label{eq:BP}
\frac{\Delta U}{U_\infty} = & \underbrace{
\left( 1-\sqrt{1-\frac{c_T}{8\left(k_{BP}\cdot x/D+0.2\sqrt{\beta}\right)^2}}\right)}_{\text{centerline velocity deficit}
}\\
& \cdot 
\underbrace{
\exp \left(-\frac{\left((z-z_h)/D\right)^2+\left(y/D\right)^2}{2\left(k_{BP}\cdot x/D+0.2\sqrt{\beta}\right)^2}\right)}_{\text{Gaussian velocity profile; 0 for $(z-z_h)=y=0$}
}
\end{align}
with
\begin{equation}\label{eq:BP_beta}
\beta = \frac{1+\sqrt{1-c_T}}{2\sqrt{1-c_T}}.
\end{equation}
$k_{BP}$ is a tunable parameter, $z_h$ denotes the hub height of the turbine and $y$ and $z$ denote the span-wise and wall-normal coordinates, respectively.\\
The model was modified later to also describe the wake of a yawed wind turbine \cite{bastankhah_porte-agel_2016}.

\subsection{Aitken wake model}
With a simple power law approach applied to various data sets, Aitken et al. derived the following relation \cite{Aitken2014, porteagel2020wind} for the normalized centerline velocity deficit,
\begin{equation}
    \frac{\Delta U}{U_\infty} = 0.56\left(\frac{x}{D}\right)^{-0.57}.
\end{equation}
Note that the exponent found by Aitken et al., $-0.57$, is close to the one predicted in equation \ref{eq:EQ_U}, $-2/3$.\\
Since this model does not have any tunable parameter, we modify it to adapt for the different wake growth rates with the parameter $A$,
\begin{equation}
    \frac{\Delta U}{U_\infty} = 0.56\left(\frac{x/D}{A}\right)^{-0.57}.
\end{equation}

\subsection{Frandsen wake model}
Similar to the Jensen wake model, the Frandsen wake model assumes a top-hat shaped velocity deficit in the wake \cite{Frandsen2006}. It is derived using conservation of mass and momentum, and we use the simplified version suggested by the authors in agreement with their findings for an infinite row of two-dimensional obstacles, where the normalized velocity deficit is given by
\begin{equation}
    \frac{\Delta U}{U_\infty}=\frac{1}{2}\left(1-\sqrt{1-\frac{2c_T}{\beta + k_F x/D}}\right), 
\end{equation}
where $\beta$ is given by equation \ref{eq:BP_beta}.
For small $c_T$, the authors state that $k_F=\mathcal{O}(10\cdot k_j)$.


\section{Additional wind tunnel data}\label{secA2b}
\begin{figure}
    \begin{minipage}[t]{.49\textwidth}
        \centering
        \stackinset{l}{0.1cm}{t}{0.2cm}{\parbox{0.5cm}{(a)}}{
        \includegraphics[width=\textwidth]{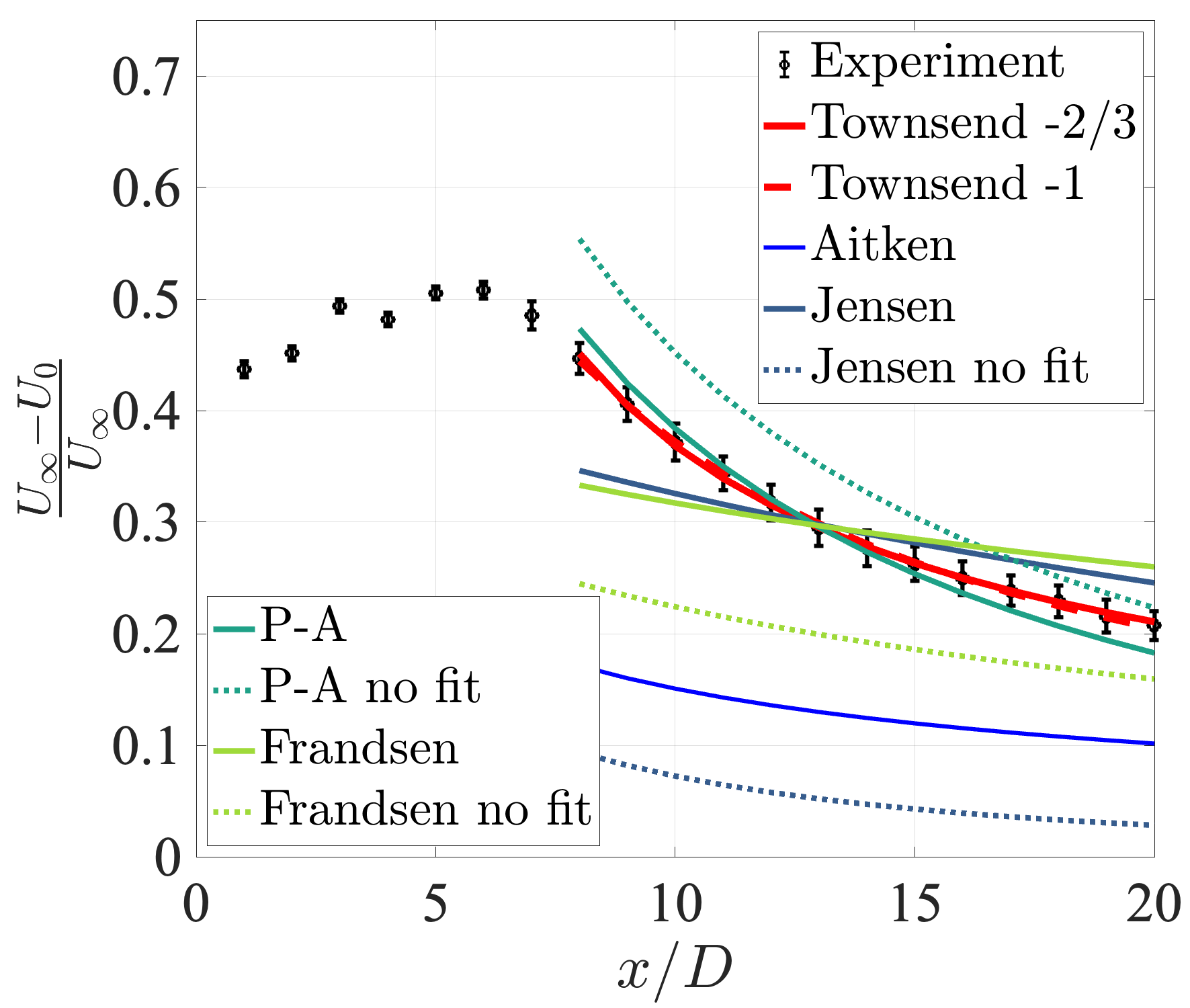}
        }
    \end{minipage}
    \hfill
    \begin{minipage}[t]{.49\textwidth}
        \centering
        \stackinset{l}{0.1cm}{t}{0.2cm}{\parbox{0.5cm}{(b)}}{
        \includegraphics[width=\textwidth]{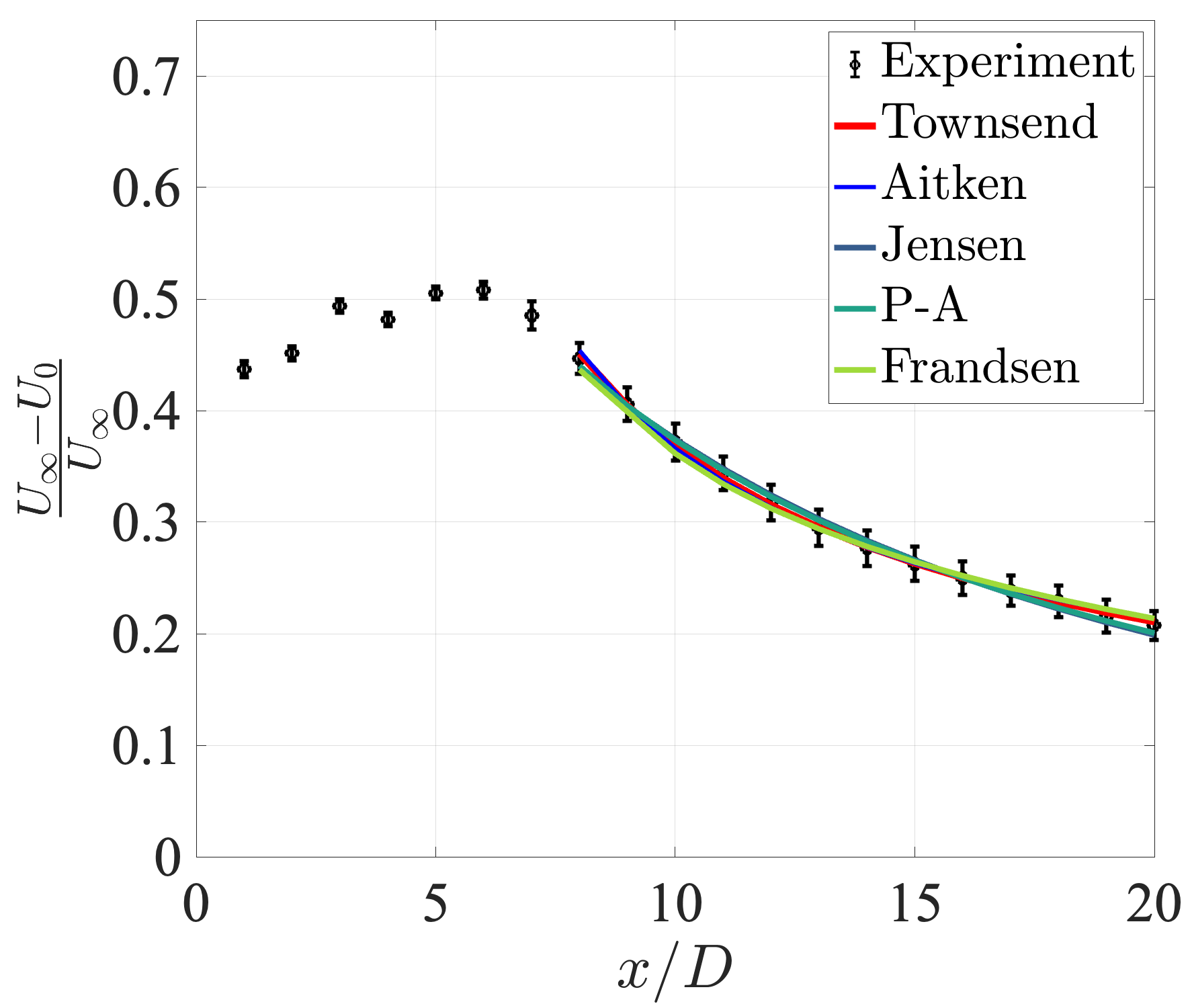}
        }
    \end{minipage}  
    \label{fig:stream}
    \caption{Downstream evolution of the normalized mean velocity deficit. Turbulent inflow without virtual origin (a), turbulent inflow with virtual origin (b)}
\end{figure}
\unskip
\pagebreak
\section{Additional field data}\label{secA3}
\begin{figure}[H]
    \begin{minipage}[t]{.49\textwidth}
        \centering
        \stackinset{l}{0.1cm}{t}{0.2cm}{\parbox{0.5cm}{(a)}}{
        \includegraphics[width=\textwidth]{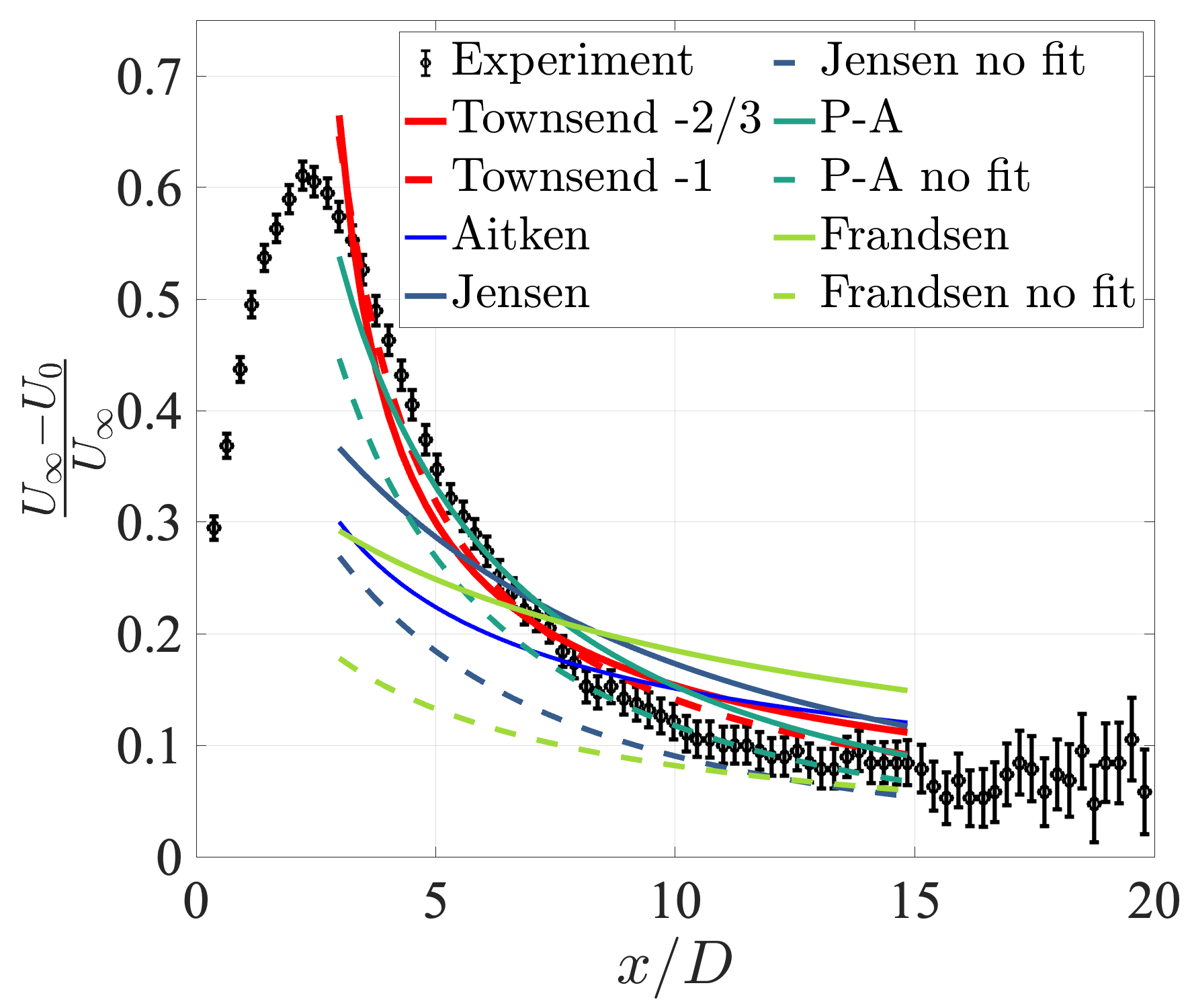}
        }
    \end{minipage}
    \hfill
    \begin{minipage}[t]{.49\textwidth}
        \centering
        \stackinset{l}{0.1cm}{t}{0.2cm}{\parbox{0.5cm}{(b)}}{
        \includegraphics[width=\textwidth]{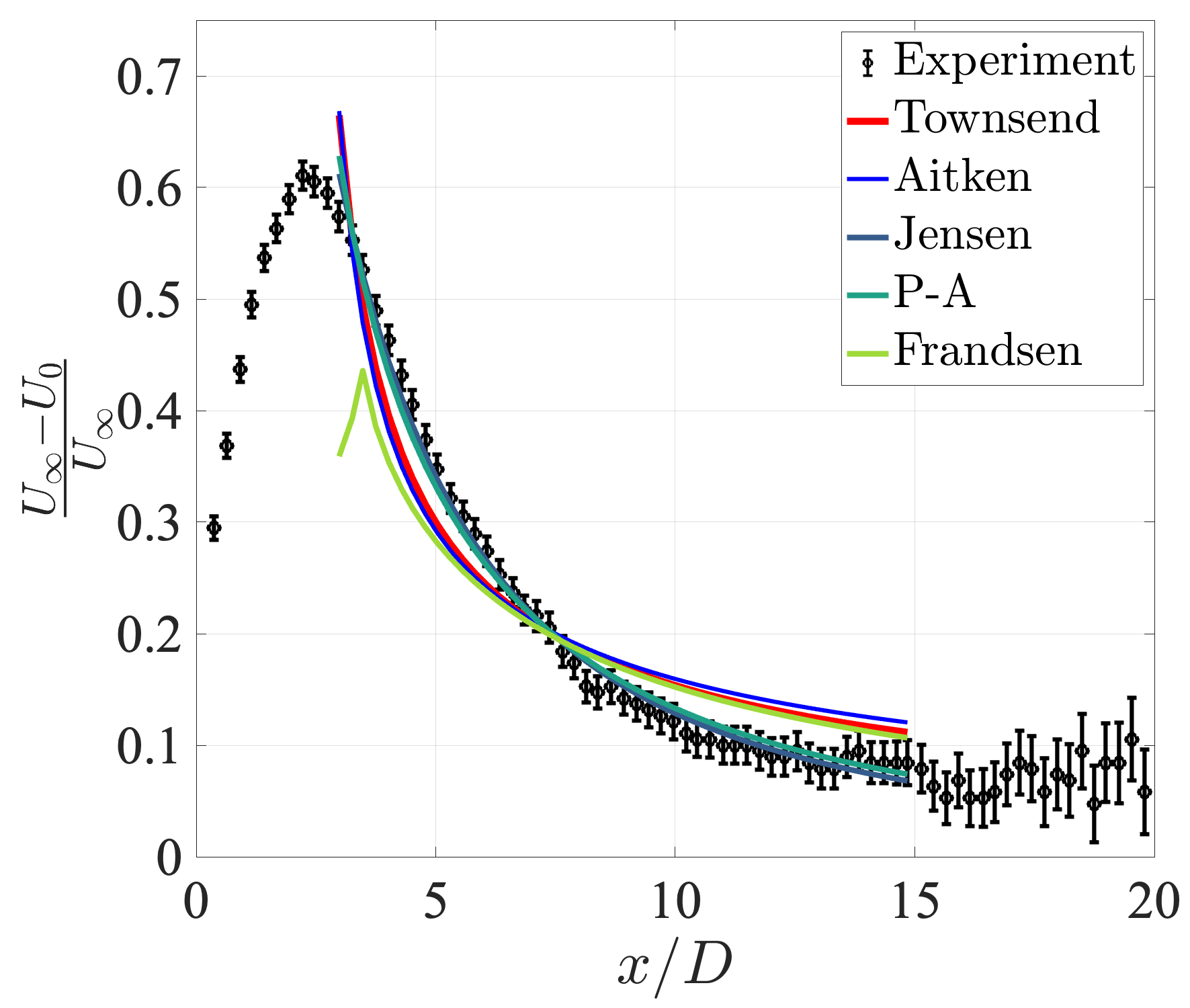}
        }
    \end{minipage}  
    \label{fig:stream}
    \caption{Downstream evolution of the normalized velocity deficit: Data measured at FINO 1 by a nacelle-mounted LiDAR with a velocity range of $8-10\:\text{ms}^{-1}$, extracted from \cite{Gallacher2014} with $c_T$ values from \cite{Soesanto2022}. Error bars indicate the standard deviation of the average over all velocity measurements, normalized by the square root of the number of measurements. Engineering wake models fitted adapting $k$ and (a) no virtual origin; (b) virtual origin.}
\end{figure}
\unskip

\begin{figure}[H]
    \begin{minipage}[t]{.49\textwidth}
        \centering
        \stackinset{l}{0.1cm}{t}{0.2cm}{\parbox{0.5cm}{(a)}}{
        \includegraphics[width=\textwidth]{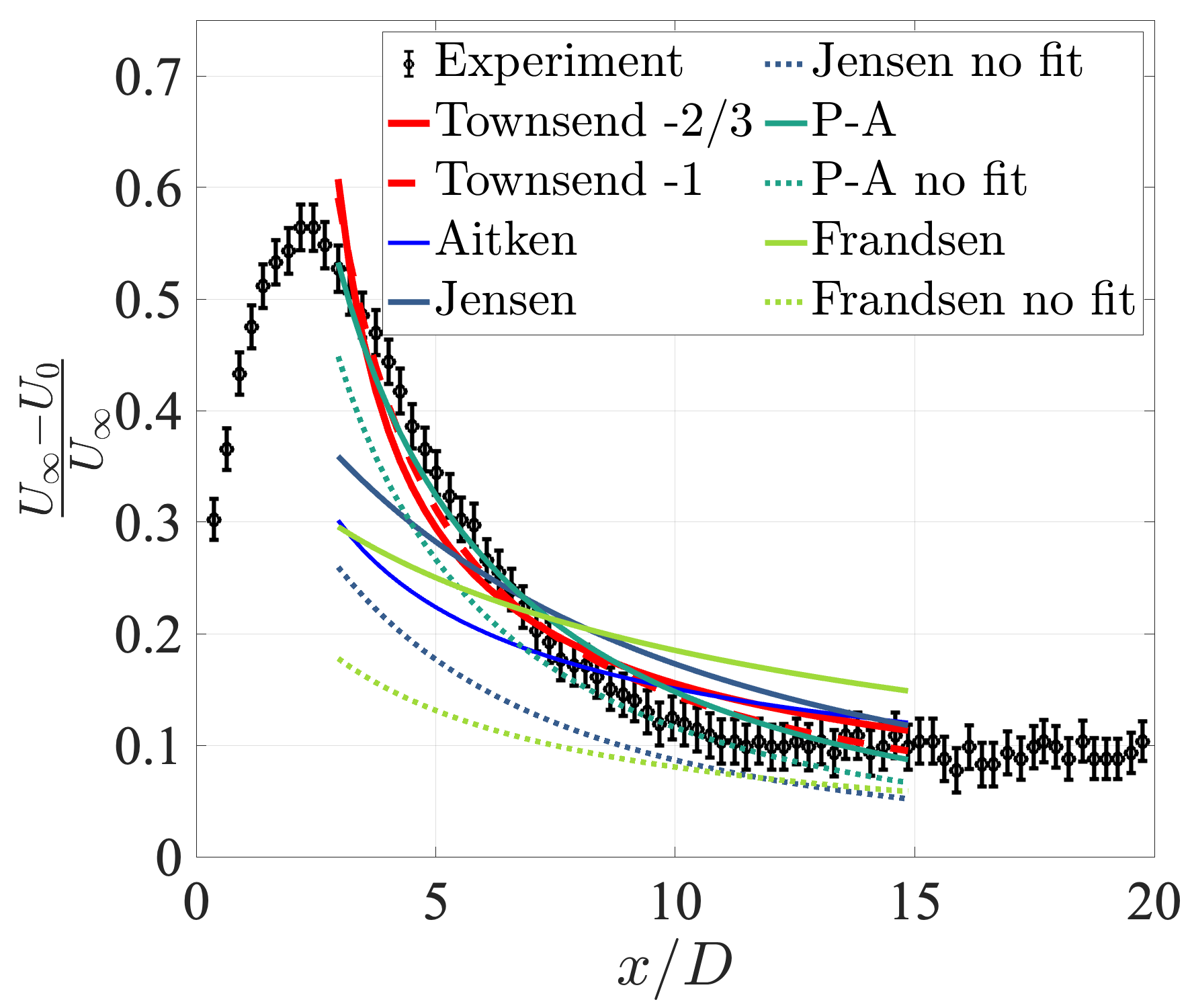}
        }
    \end{minipage}
    \hfill
    \begin{minipage}[t]{.49\textwidth}
        \centering
        \stackinset{l}{0.1cm}{t}{0.2cm}{\parbox{0.5cm}{(b)}}{
        \includegraphics[width=\textwidth]{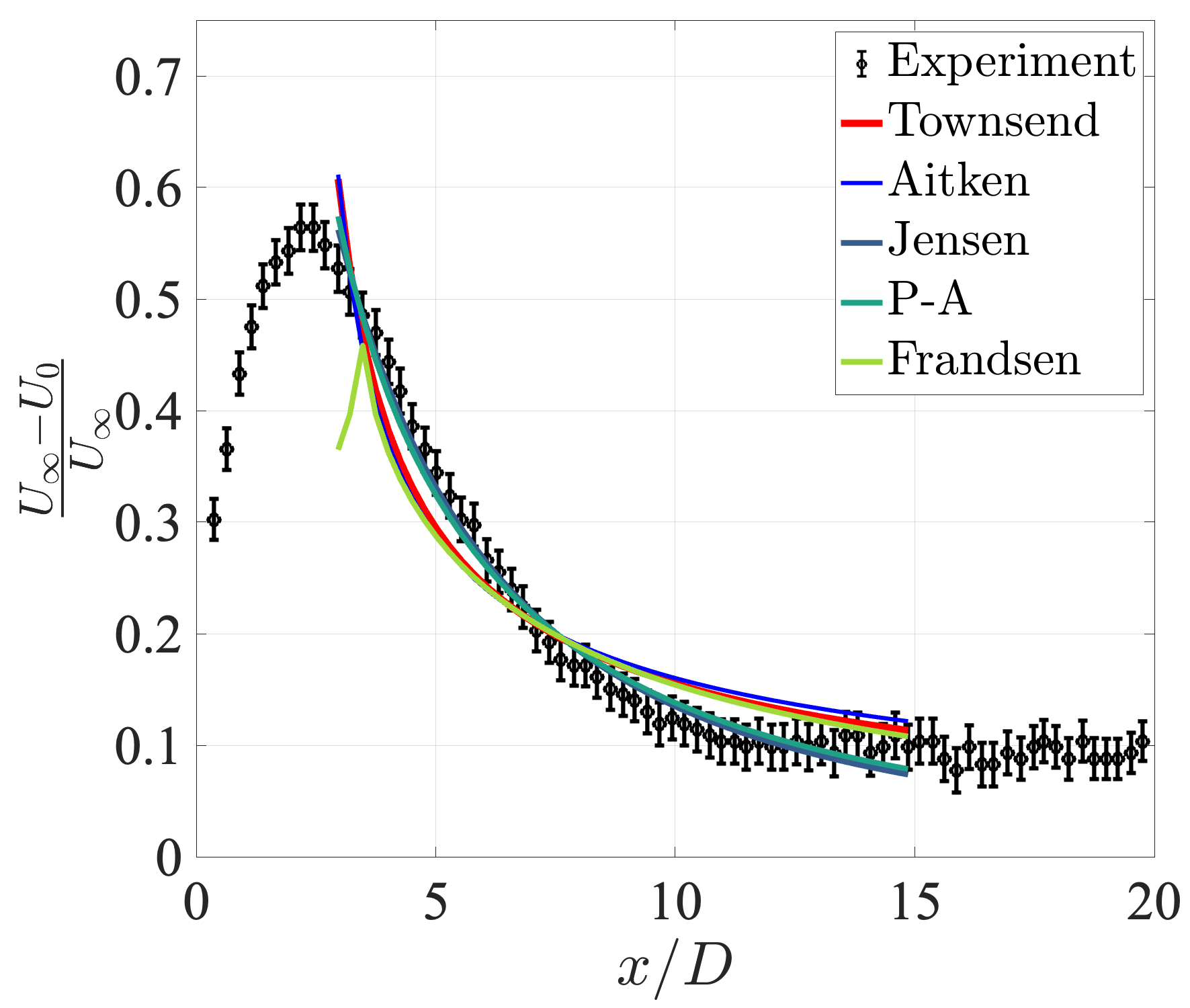}
        }
    \end{minipage}  
    \label{fig:stream}
    \caption{Downstream evolution of the normalized velocity deficit: Data measured at FINO 1 by a nacelle-mounted LiDAR with a velocity range of $10-12\:\text{ms}^{-1}$, extracted from \cite{Gallacher2014} with $c_T$ values from \cite{Soesanto2022}. Error bars indicate the standard deviation of the average over all velocity measurements, normalized by the square root of the number of measurements. Engineering wake models fitted adapting $k$ and (a) no virtual origin; (b) virtual origin.}
\end{figure}
\unskip

\begin{figure}[H]
    \begin{minipage}[t]{.49\textwidth}
        \centering
        \stackinset{l}{0.1cm}{t}{0.2cm}{\parbox{0.5cm}{(a)}}{
        \includegraphics[width=\textwidth]{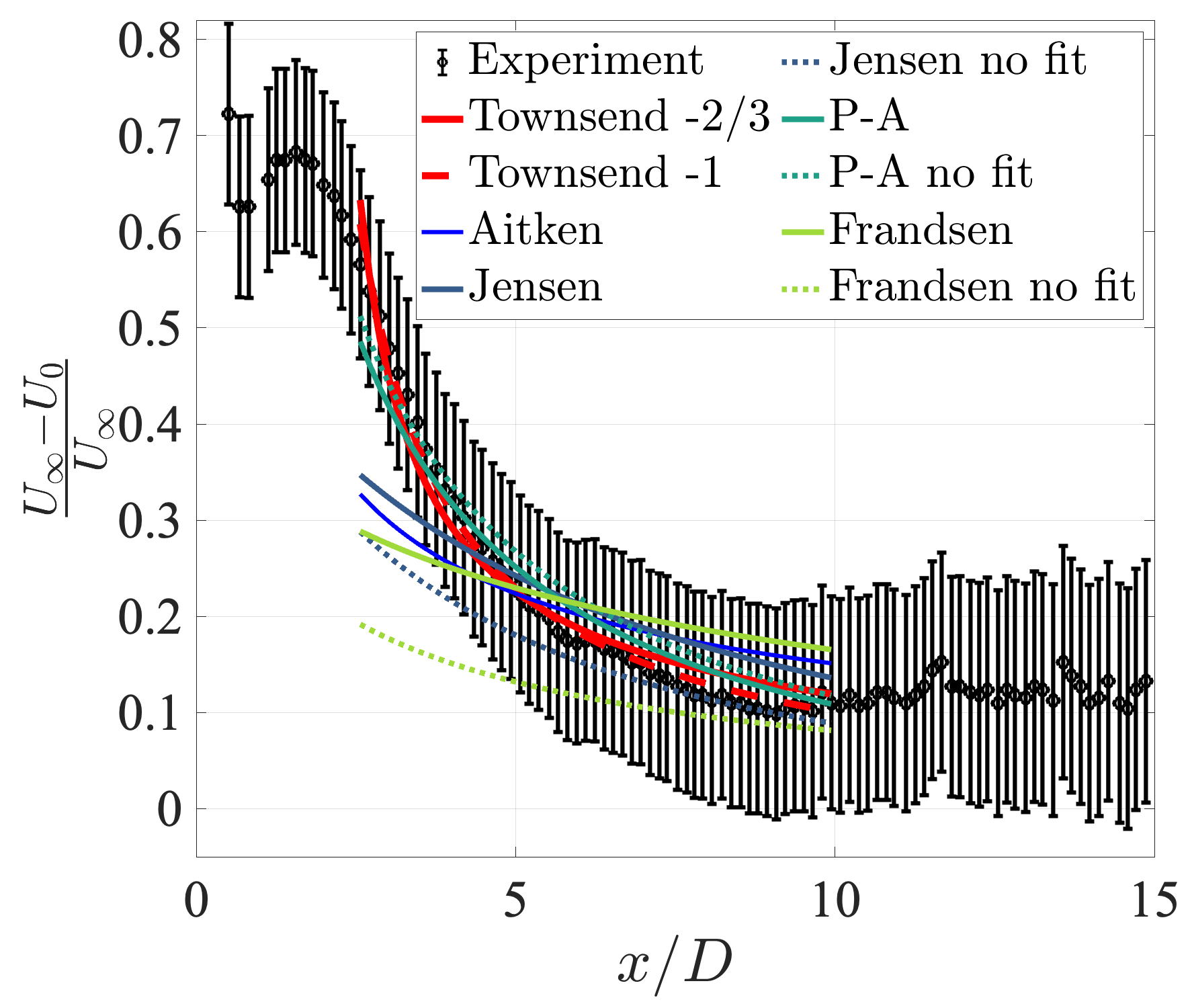}
        }
    \end{minipage}
    \hfill
    \begin{minipage}[t]{.49\textwidth}
        \centering
        \stackinset{l}{0.1cm}{t}{0.2cm}{\parbox{0.5cm}{(b)}}{
        \includegraphics[width=\textwidth]{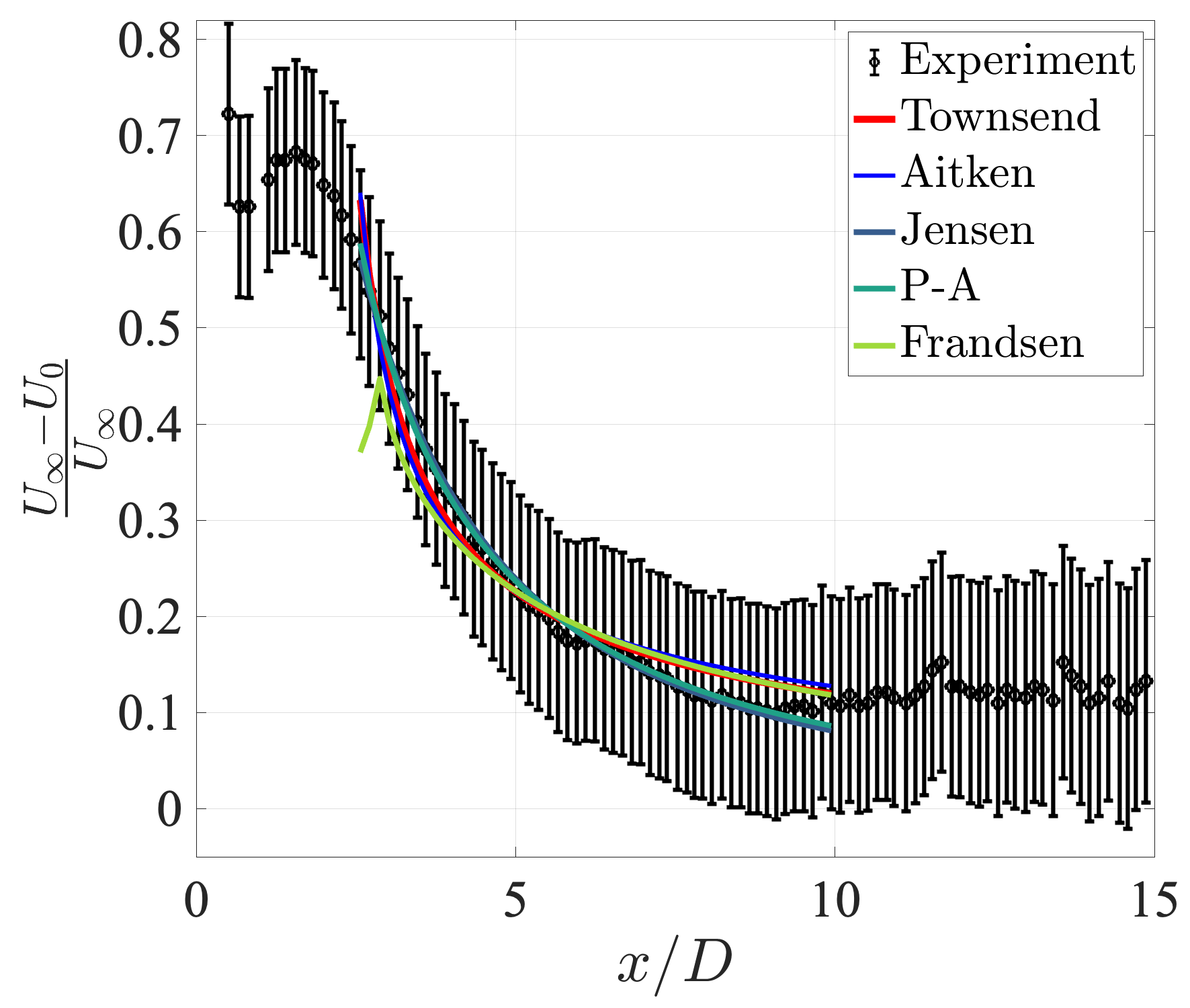}
        }
    \end{minipage}  
    \label{fig:stream}
    \caption{Downstream evolution of the normalized velocity deficit: Data measured at the FINO 1 platform by LiDAR, extracted from \cite{KRISHNAMURTHY2017428}. Error bars indicate the standard deviation of the average over all velocity measurements. Engineering wake models fitted adapting $k$ and (a) no virtual origin; (b) virtual origin.}
\end{figure}
\unskip

\begin{figure}[H]
    \begin{minipage}[t]{.49\textwidth}
        \centering
        \stackinset{l}{0.1cm}{t}{0.2cm}{\parbox{0.5cm}{(a)}}{
        \includegraphics[width=\textwidth]{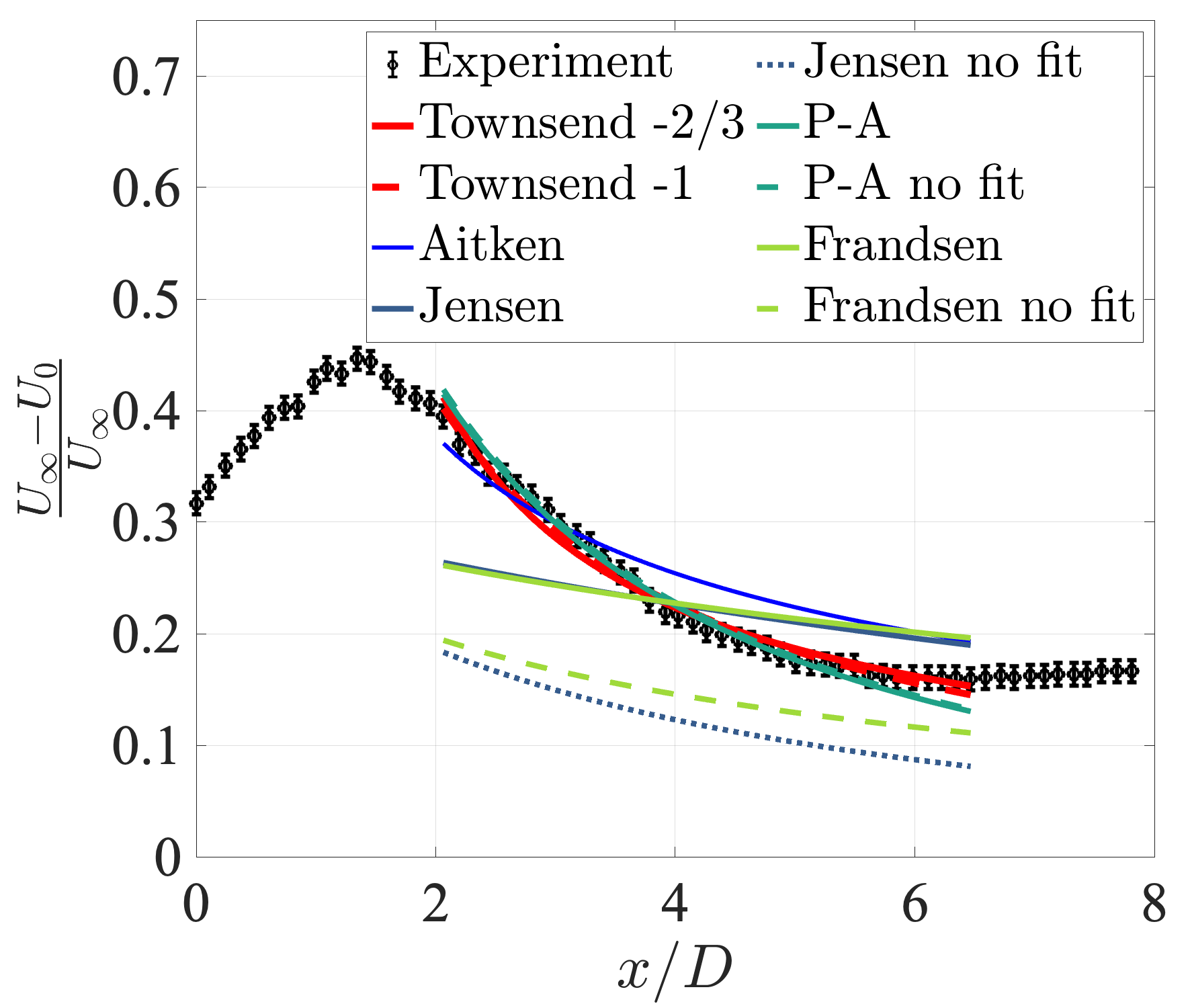}
        }
    \end{minipage}
    \hfill
    \begin{minipage}[t]{.49\textwidth}
        \centering
        \stackinset{l}{0.1cm}{t}{0.2cm}{\parbox{0.5cm}{(b)}}{
        \includegraphics[width=\textwidth]{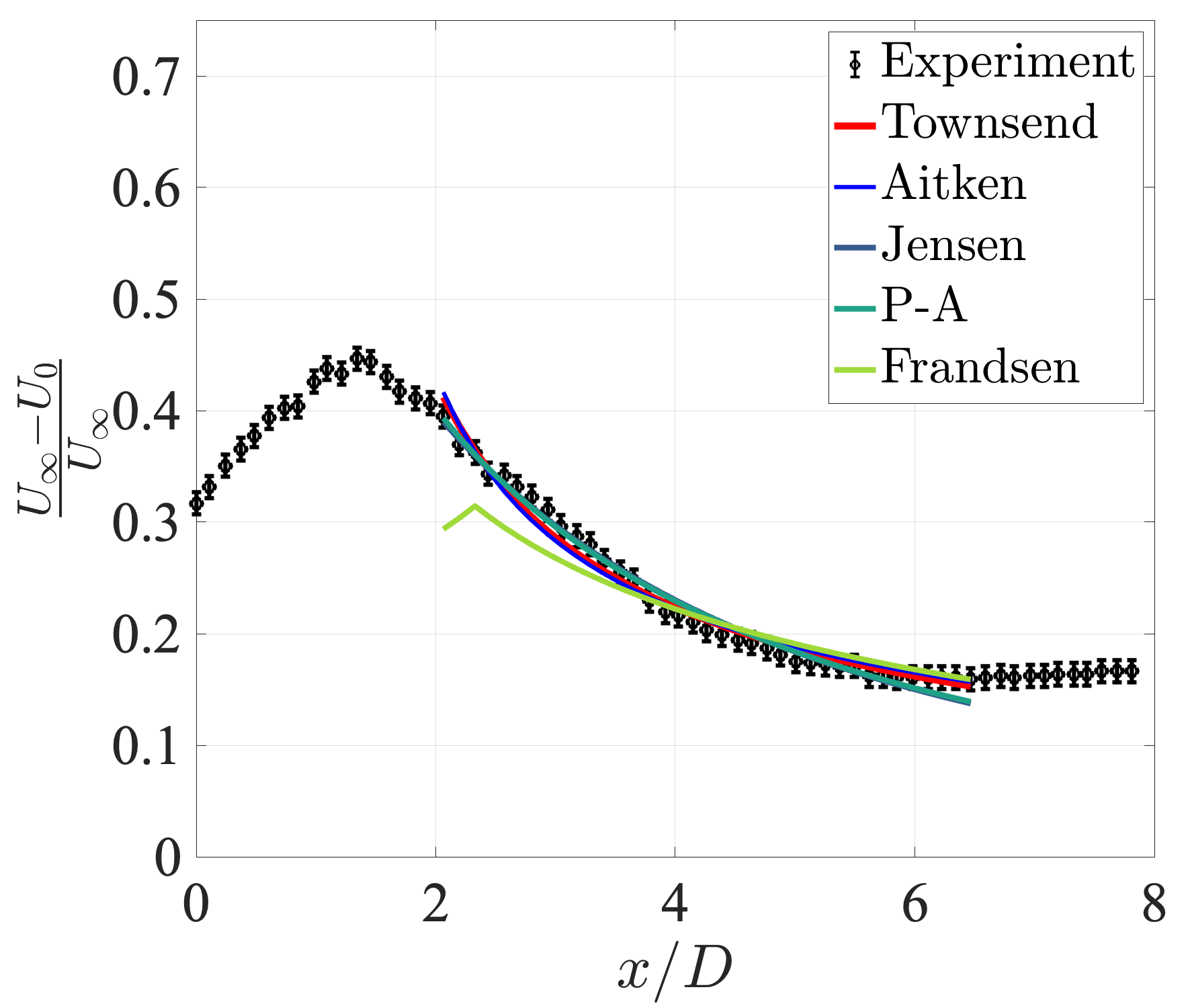}
        }
    \end{minipage}  
    \label{fig:stream}
    \caption{Downstream evolution of the normalized velocity deficit: Data measured onshore by LiDAR within the SMARTEOLE project, extracted from \cite{HEGAZY2022457}; the wake slightly interacts with the wake of a downstream turbine. Error bars are indicating variations of 0.01 of the normalized velocity deficit. Engineering wake models fitted adapting $k$ and (a) no virtual origin; (b) virtual origin.}
\end{figure}
\unskip

\begin{figure}[H]
    \begin{minipage}[t]{.49\textwidth}
        \centering
        \stackinset{l}{0.1cm}{t}{0.2cm}{\parbox{0.5cm}{(a)}}{
        \includegraphics[width=\textwidth]{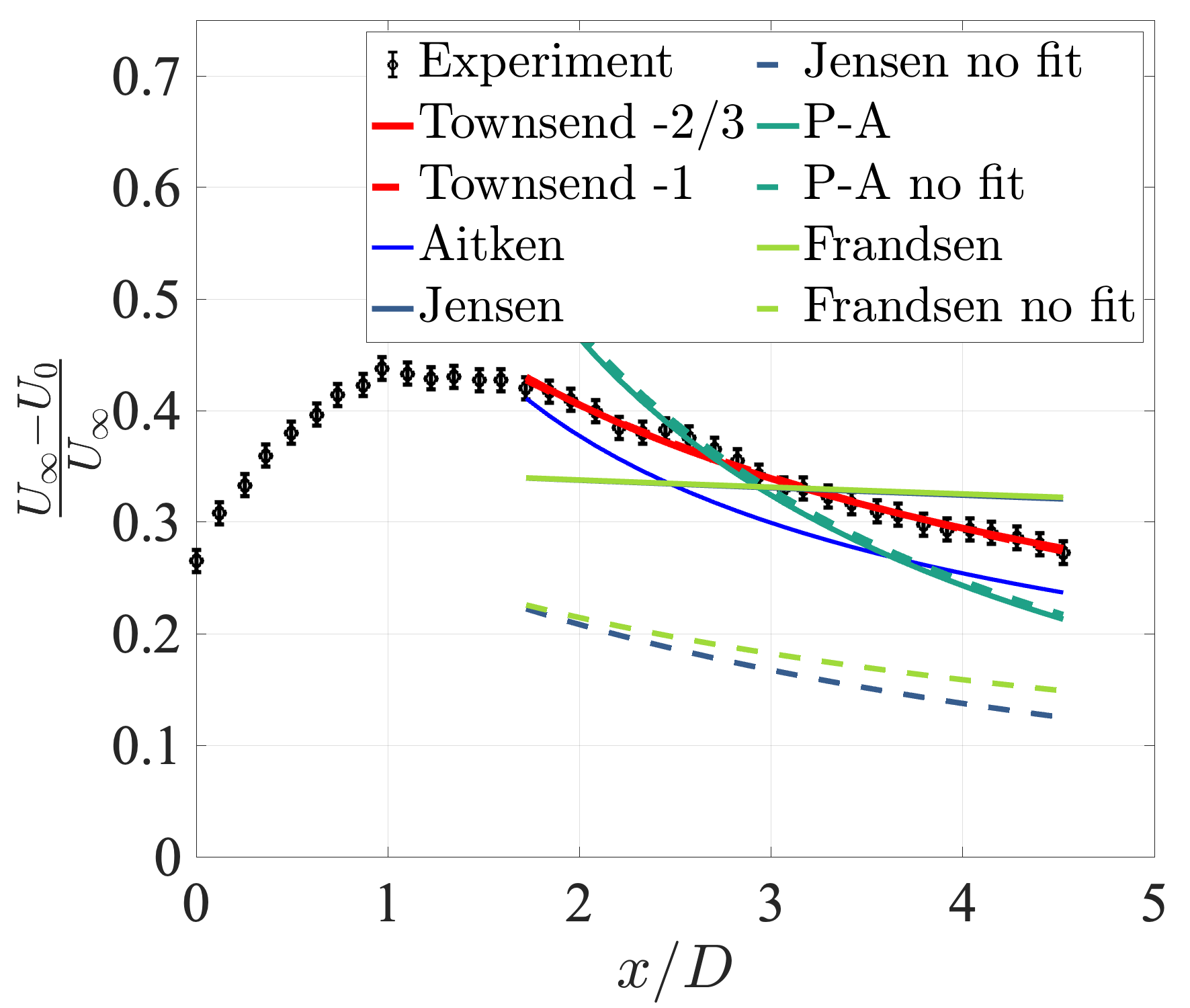}
        }
    \end{minipage}
    \hfill
    \begin{minipage}[t]{.49\textwidth}
        \centering
        \stackinset{l}{0.1cm}{t}{0.2cm}{\parbox{0.5cm}{(b)}}{
        \includegraphics[width=\textwidth]{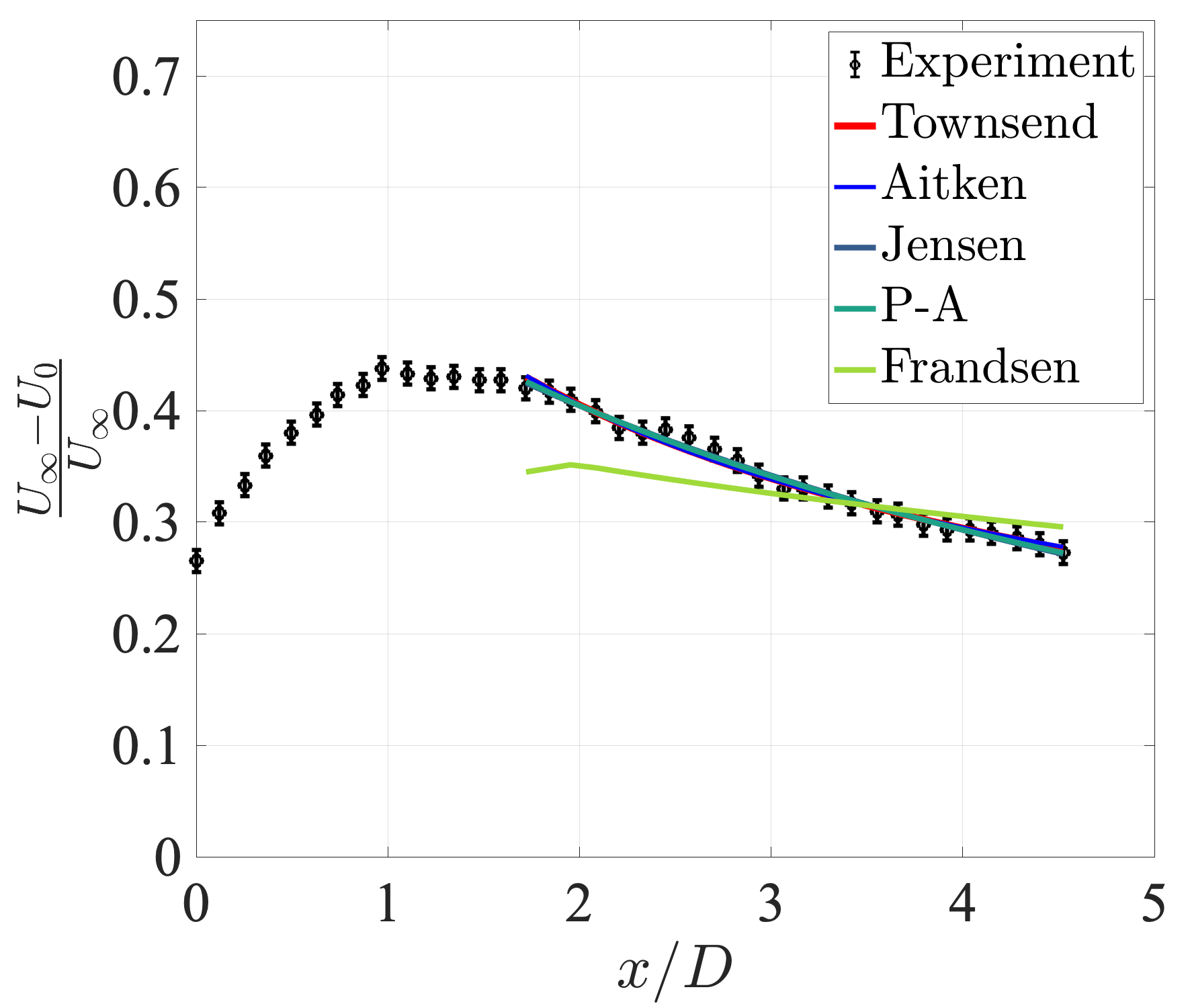}
        }
    \end{minipage}  
    \label{fig:stream}
    \caption{Downstream evolution of the normalized velocity deficit: Data measured onshore by LiDAR within the SMARTEOLE project, extracted from \cite{HEGAZY2022457}; the turbine slightly interacts with the wake of an upstream turbine. Error bars are indicating variations of 0.01 of the normalized velocity deficit. Engineering wake models fitted adapting $k$ and (a) no virtual origin; (b) virtual origin.}
\end{figure}
\unskip

\begin{figure}[H]
    \begin{minipage}[t]{.49\textwidth}
        \centering
        \stackinset{l}{0.1cm}{t}{0.2cm}{\parbox{0.5cm}{(a)}}{
        \includegraphics[width=\textwidth]{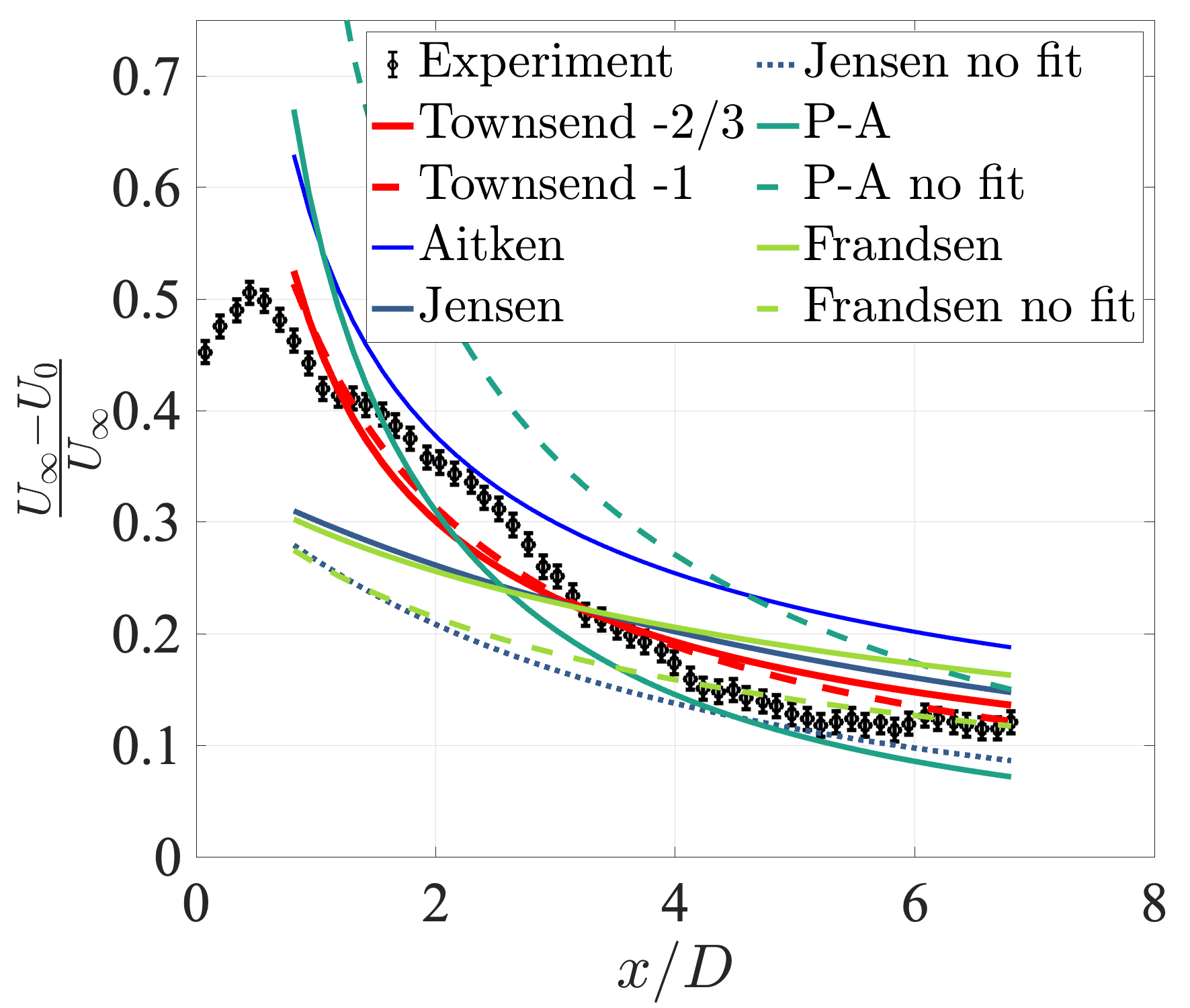}
        }
    \end{minipage}
    \hfill
    \begin{minipage}[t]{.49\textwidth}
        \centering
        \stackinset{l}{0.1cm}{t}{0.2cm}{\parbox{0.5cm}{(b)}}{
        \includegraphics[width=\textwidth]{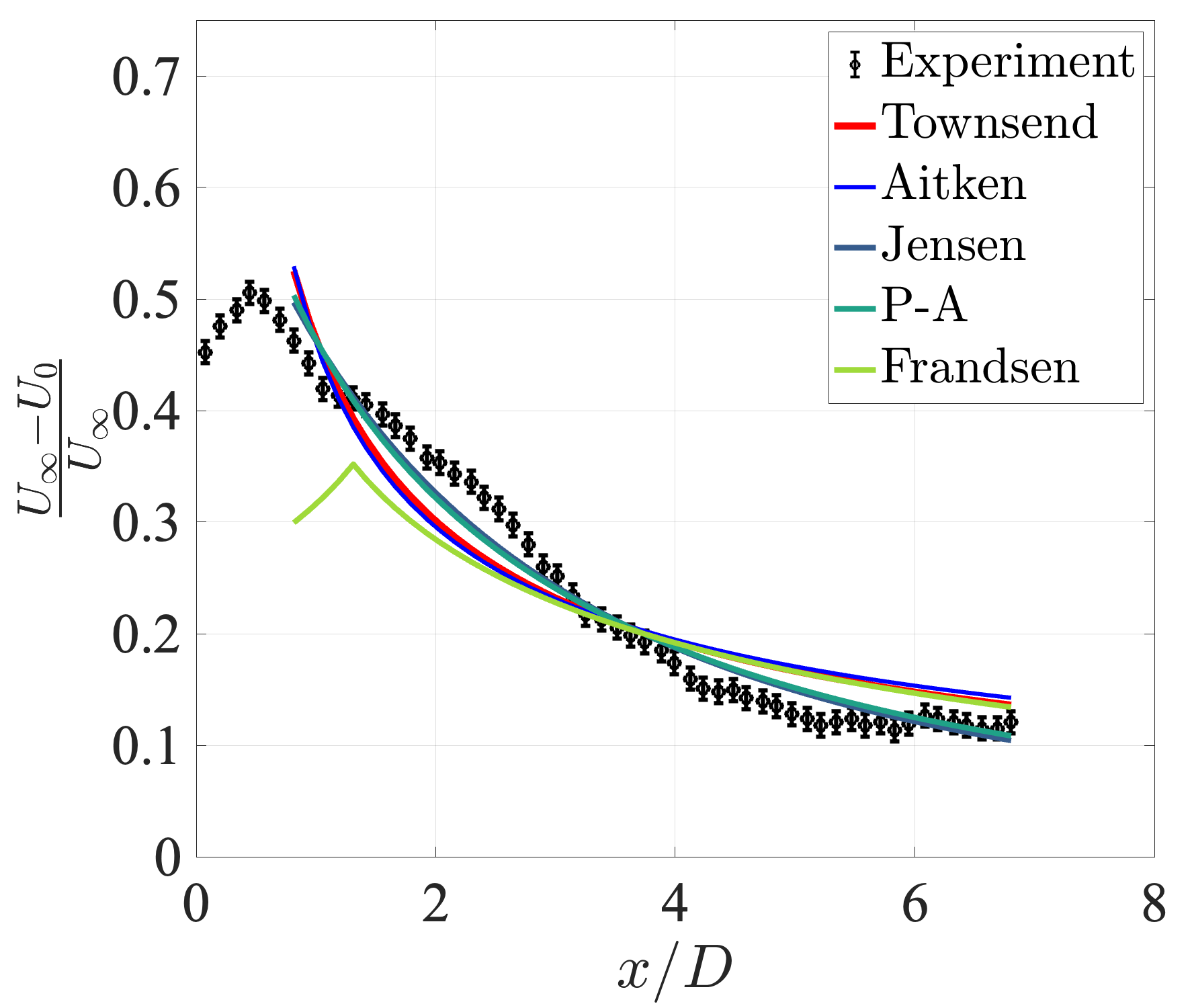}
        }
    \end{minipage}  
    \label{fig:stream}
    \caption{Downstream evolution of the normalized velocity deficit: Data measured onshore by LiDAR within the SMARTEOLE project, extracted from \cite{HEGAZY2022457}. Error bars are indicating variations of 0.01 of the normalized velocity deficit. Engineering wake models fitted adapting $k$ and (a) no virtual origin; (b) virtual origin.}
\end{figure}
\unskip

\begin{figure}[H]
    \begin{minipage}[t]{.49\textwidth}
        \centering
        \stackinset{l}{0.1cm}{t}{0.2cm}{\parbox{0.5cm}{(a)}}{
        \includegraphics[width=\textwidth]{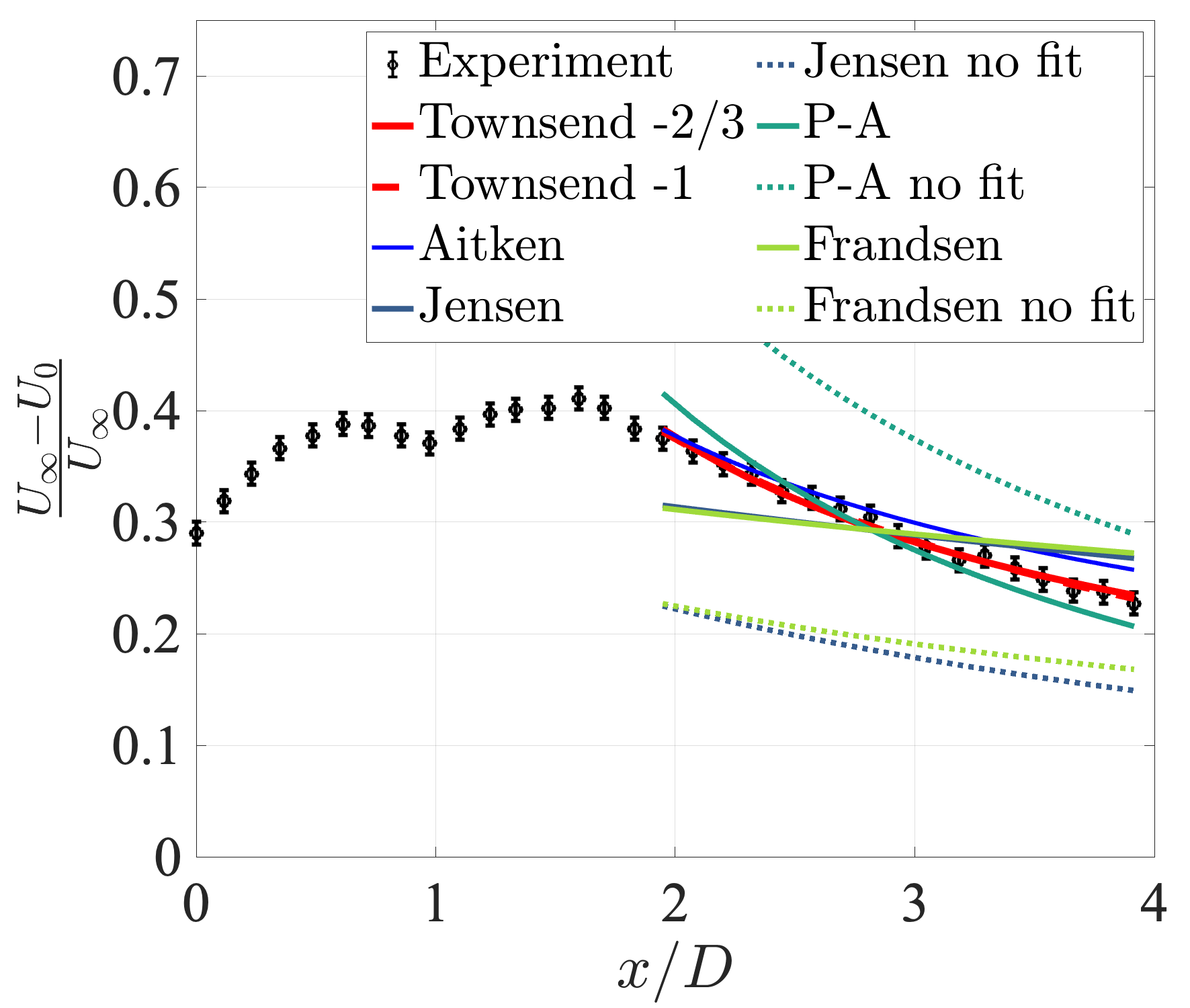}
        }
    \end{minipage}
    \hfill
    \begin{minipage}[t]{.49\textwidth}
        \centering
        \stackinset{l}{0.1cm}{t}{0.2cm}{\parbox{0.5cm}{(b)}}{
        \includegraphics[width=\textwidth]{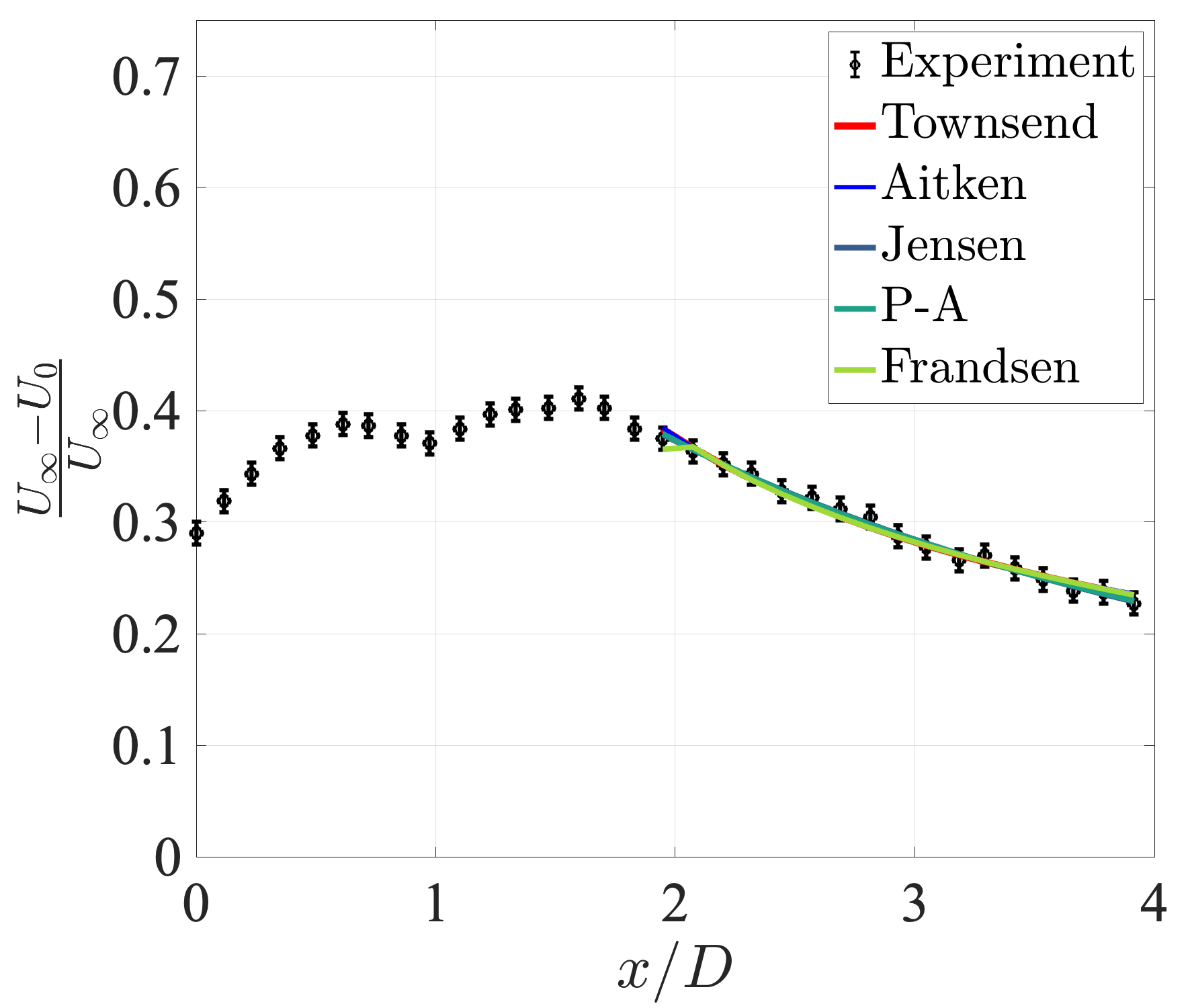}
        }
    \end{minipage}  
    \label{fig:stream}
    \caption{Downstream evolution of the normalized velocity deficit: Data measured onshore by LiDAR within the SMARTEOLE project, extracted from \cite{HEGAZY2022457}. Error bars are indicating variations of 0.01 of the normalized velocity deficit. Engineering wake models fitted adapting $k$ and (a) no virtual origin; (b) virtual origin.}
\end{figure}
\unskip

\pagebreak
\section{Fitting only a virtual origin}\label{secA4}
In this section, we present the results from adding a virtual origin as fitting parameter to the engineering models while using theoretical values for the wake expansion rates $k$. The values used for $k$ and obtained for $x_0$ can be found in table \ref{tab1a}. Note that, while we aimed at keeping $k$ close to the theoretical values, we adapted them to give reasonable results for the fits.
\begin{figure}[H]
    \begin{minipage}[t]{.49\textwidth}
        \centering
        \stackinset{l}{0.1cm}{t}{0.2cm}{\parbox{0.5cm}{(a)}}{
        \includegraphics[width=\textwidth]{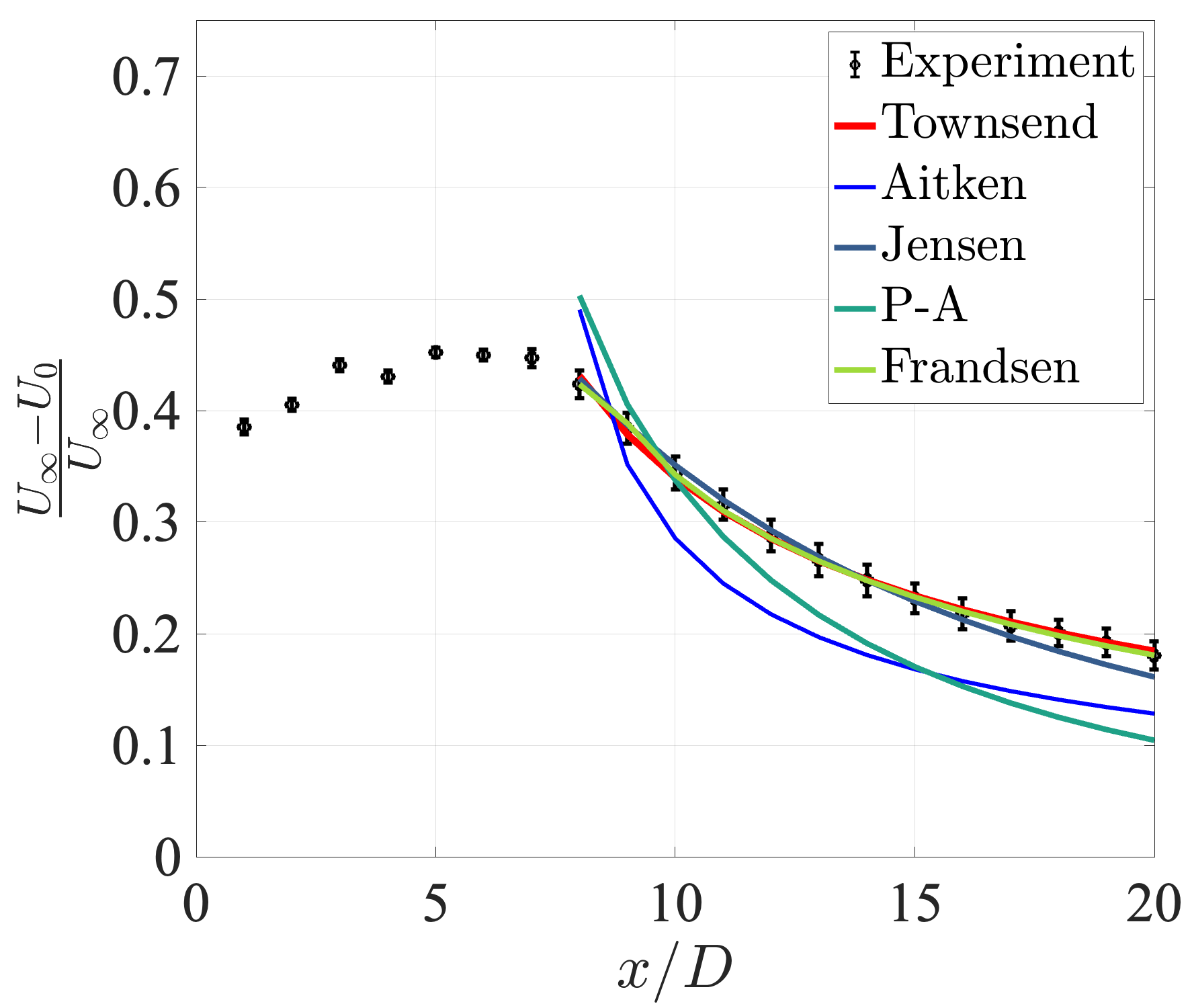}
        }
    \end{minipage}
    \hfill
    \begin{minipage}[t]{.49\textwidth}
        \centering
        \stackinset{l}{0.1cm}{t}{0.2cm}{\parbox{0.5cm}{(b)}}{
        \includegraphics[width=\textwidth]{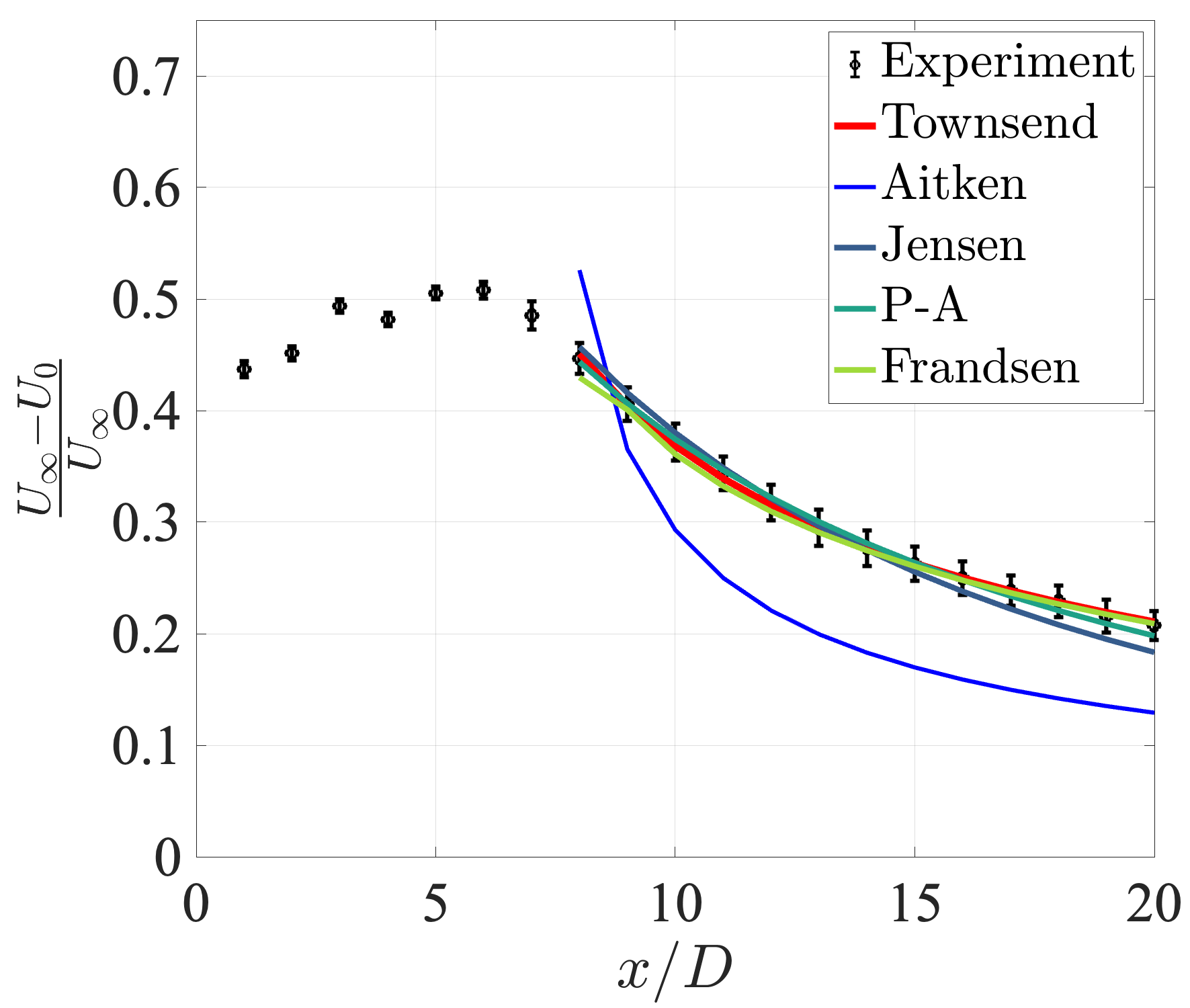}
        }
    \end{minipage}  
    \label{fig:Lab}
    \caption{Downstream evolution of the normalized velocity deficit. Only a virtual origin is fitted for the engineering wake models. (a) laminar inflow; (b) turbulent inflow.}
\end{figure}
\unskip

\begin{figure}[H]
        \centering
        \includegraphics[width=\textwidth]{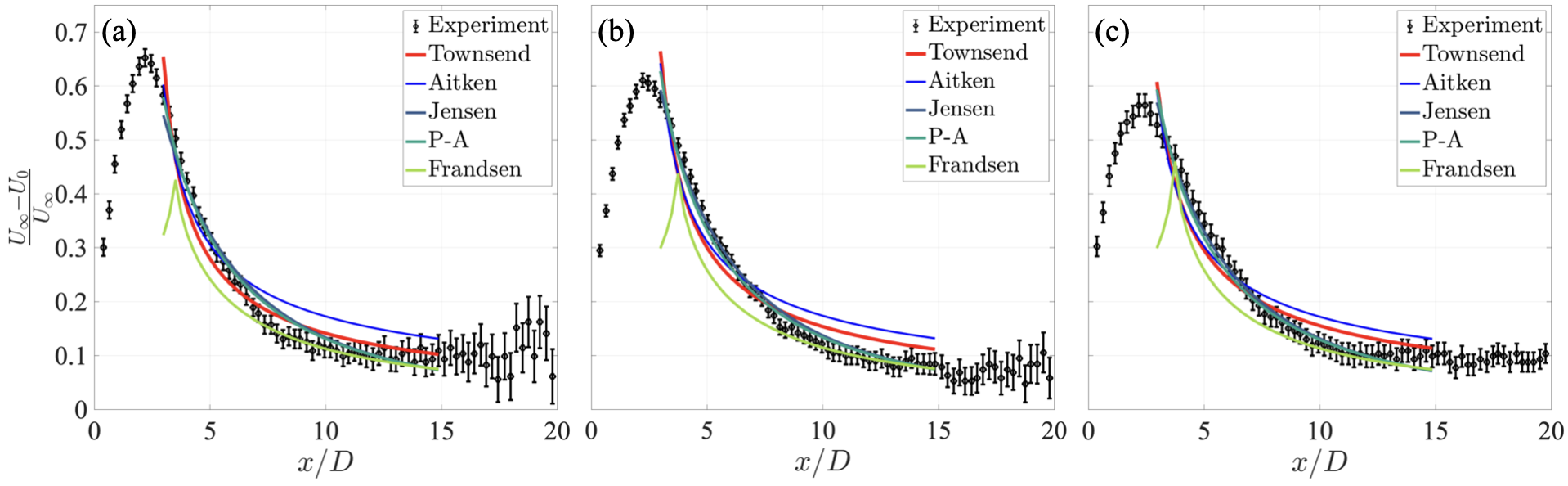}
    \label{fig:VOGallacher}
    \caption{Downstream evolution of the normalized velocity deficit: Data measured at FINO 1 by a nacelle-mounted LiDAR, extracted from \cite{Gallacher2014} with $c_T$ values from \cite{Soesanto2022}. Only a virtual origin is fitted for the engineering wake models. Velocity range of (a) $6-8\:\text{ms}^{-1}$; (b) $8-10\:\text{ms}^{-1}$; (c) $10-12\:\text{ms}^{-1}$.}
\end{figure}
\unskip

\begin{figure}[H]
        \centering
        \includegraphics[width=.5\textwidth]{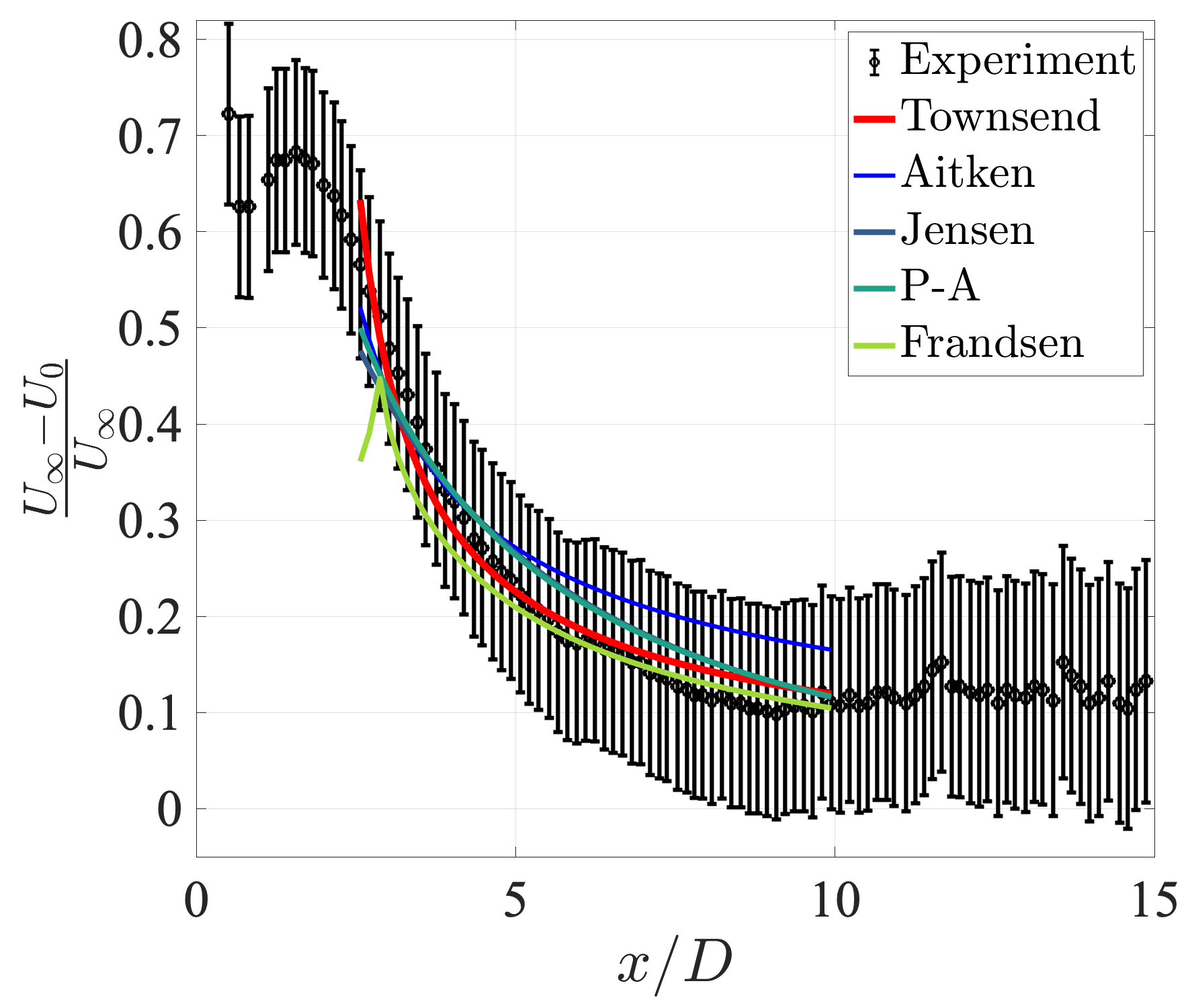}
    \label{fig:VOKrish}
    \caption{Downstream evolution of the normalized velocity deficit: Data measured at the FINO 1 platform by LiDAR, extracted from \cite{KRISHNAMURTHY2017428}. Only a virtual origin is fitted for the engineering wake models.}
\end{figure}
\unskip

\begin{figure}[H]
        \centering
        \includegraphics[width=\textwidth]{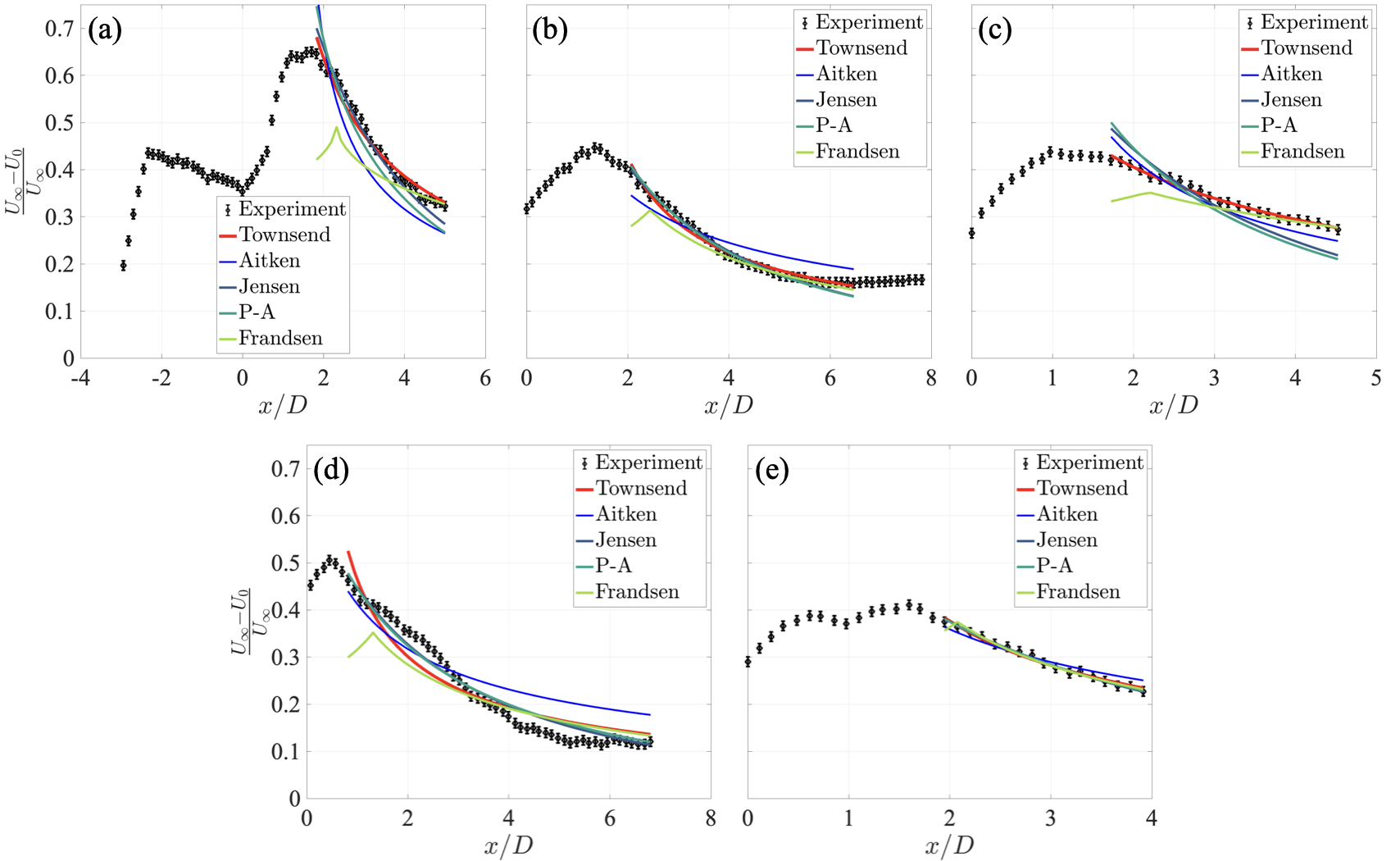}
    \label{fig:VOSmarteole}
    \caption{Downstream evolution of the normalized velocity deficit: Data measured onshore by LiDAR within the SMARTEOLE project, extracted from \cite{HEGAZY2022457}. Error bars are indicating variations of 0.01 of the normalized velocity deficit. Only a virtual origin is fitted for the engineering wake models. (a) wake of a turbine exposed to the wake of an upstream turbine; (b) the wake interacts slightly with the wake of a downstream turbine; (c) wake of a turbine partially exposed to the wake of an upstream turbine; (d) wake of a wind turbine; (e) wake of a wind turbine.}
\end{figure}
\unskip
